\documentclass[12pt]{iopart}
\usepackage{hyperref}
\usepackage{iopams}
\usepackage{setstack}
\usepackage{graphicx}
\usepackage{subfig}
\begin{document}
	\title[Evolution of Flat Band and Van Hove Singularities with Interlayer Coupling in TBG]{Evolution of Flat Band and Van Hove Singularities with Interlayer Coupling in Twisted Bilayer Graphene}
	\author{Veerpal$^1$ and B.R.K. Nanda$^2$ and Ajay$^3$}
	\address{ Department of Physics, Indian Institute of Technology Roorkee, India$^{1,2}$ and Department of Physics, Indian Institute of Technology Madras, Chennai, India$^3$}
	\ead{veerpal@ph.iitr.ac.in$^1$, nandab@smail.iitm.ac.in$^2$ and ajay@ph.iitr.ac.in$^3$}
	\begin{abstract}
     Here we present a theoretical analysis (applicable to all twist angles of TBG) of band dispersion and density of states in TBG relating evolution of flat band and Van-Hove singularities with evolution of interlayer coupling in TBG. A simple tight binding Hamiltonian with environment dependent interlayer hopping and incorporated with internal configuration of carbon atoms inside a supercell is used to calculate band dispersion and density of states in TBG. Various Hamiltonian parameters and functional form of interlayer hopping applicable to a wide range of twist angles in TBG is estimated by fitting calculated dispersion and density of states with available experimentally observed dispersion and density of states in Graphene, AB-stacked bilayer graphene and some TBG systems. Computationally obtained band dispersion reveal that flat band in TBG occurs very close to Dirac point of graphene and only along linear dimension of two-dimensional wave vector space connecting two closest Dirac points of two graphene layers of TBG. Computed dispersion and density of states for TBG show that instead of being restricted to TBG with discrete set of magic angles, flat band and Van-Hove singularities appear for all TBG systems having twist angle up to $\approx 3.5^{\circ}$ as observed experimentally.
		
	\end{abstract}
\maketitle	

	\section{Introduction}\label{sec1}
The thought that graphene based layered materials with a relative twist between the graphene layers can offer tunability of electronic properties driven by twist angle motivated scientific community to investigate these materials\cite{2007-Santos,2008-Shallcross,2010-Shallcross,2012-Santos}. The observation of  Van-Hove singularities near Fermi level\cite{2010-Guohong,2011-Luican} for TBG with small twist angles and prediction of flat band near Fermi level, in magic angle twisted bilayer graphene \cite{2011-Bistritzer} caused an urge in the scientific community to explore these materials more rigorously which was further intensified by discovery of unconventional superconductivity in magic angle twisted bilayer graphene \cite{1-2018-March-Cao}. Investigation of TBG has resulted in discovery and prediction of many interesting properties in TBG, e.g., moiré superlattice \cite{2019-Alexander,2019-Yonglong,2019-Youngjoon,2019-Yuhang}, flat band near Dirac point\cite{2011-Bistritzer,2021-Lisi,2020-Utama}, Van-Hove singularities near Fermi level\cite{2010-Guohong,2019-Yonglong,2019-Alexander}, emergence of unconventional superconductivity\cite{1-2018-March-Cao,2019-Matthew,2021-Myungchul}, correlated insulator behaviour\cite{2-2018-March-Cao},anomalous Hall effect at half filling\cite{2022-Chun-Chih}, spin-orbit driven ferromagnetism\cite{2022-Jiang}, Hofstadter butterfly\cite{2022-Nadia} and many more.
Flat band in twisted bilayer graphene (TBG) persists even on room temperature while superconductivity in TBG disappears at very low critical temperature indicating that mechanism causing flat band in TBG is more sustainable in comparison to mechanism causing superconductivity. Van-Hove singularities near Fermi level are consequence of Flat band near Dirac point in TBG. We present here a theoretical model study of band dispersion and density of states in TBG based on a tight-binding model Hamiltonian equipped with environment dependent interlayer hopping, incorporated with full internal configuration of carbon atoms inside a supercell and applicable to wide range of twist angles in TBG. Various Hamiltonian parameters and functional form of interlayer hopping applicable to a wide range of twist angles in TBG is estimated by fitting calculated dispersion and density of states with experimentally observed dispersion and density of states in in Graphene, AB-stacked bilayer graphene and some TBG systems. \\
The content of following sections in this article is organized as follows. In Sec. \ref{sec2}, we briefly discuss the lattice structure of TBG. In Sec. \ref{sec3}, we present a general tight binding Hamiltonian for TBG applicable to wide range of twist angle of TBG. In Sec. \ref{sec4}, we estimate values of Hamiltonian parameters and establish the functional form of interlayer hopping for TBG applicable to wide range of twist angle of TBG. In Sec. \ref{sec5}, we present calculated energy dispersion and density of states for some TBG systems obtained after processing the Hamiltonian computationally and discuss the results.

\section{Lattice structure of TBG}	\label{sec2}
For lattice structure of TBG we follow the considerations presented in the article of reference\cite{2023-Veerpal1}, which describes the lattice structure of TBG comprehensively. Lattice structure of TBG can be considered to be modified form of either AA-stacked or AB-stacked bilayer graphene after introducing relative twist between two graphene layers of conventional bilayer graphene. This relative twist between the graphene layers creates a moire pattern in the lattice structure of TBG and each moire pattern of TBG is very specific according to the relative twist angle.
Any moire pattern of TBG can be associated with some perfectly periodic commensurate moire pattern of TBG. These moire patterns of TBG are characterized in terms of three parameters: commensurate moire period ($L_c$), corresponding minimum commensurate displacement ($\delta_c$) and corresponding commensurate twist angle ($\theta_c$). Due to perfect periodicity of commensurate moire patterns, whole moire pattern of TBG can be divided into supercells; each supercell is identical in shape, size, and internal configuration of carbon atoms. Thus, complete structure of moire pattern may be described by describing the internal configuration of one supercell. Using the considerations presented in the article of reference\cite{2023-Veerpal1}; internal configuration of carbon atoms inside a supercell of TBG can easily be simulated. Supercell of TBG corresponding to commensurate moire period equal to $L_c$ contains $N_S$ lattice points of each of $A_1$, $B_1$, $A_2$ and $B_2$ sublattices, here $N_S=L^2_c$ and $L_c$ is in units of $a_o$.
\section{Hamiltonian of the system}\label{sec3}
Experimentally observed intrinsic carrier density in graphene is $3.87\times10^{6} cm^{-2}K^{-2}\cdot T^2$ \cite{2014-Yan}, which suggests that at 300 K, intrinsic carrier density in graphene based materials would be of the order $n_{in}=3.483\times10^{11} cm^{-2}$ i.e., $0.91\times10^{-4}$ electrons per atom. 
On the basis of intrinsic carrier density it could be inferred that in graphene the $p_z$ electrons of carbon atoms are tightly bound to the carbon atoms which could occasionally participate in transport through quantum tunnelling. It has been shown in various studies that the most appropriate Hamiltonian model for graphene-based materials is tight binding model. In graphene which have triangular lattice with 2 atom basis, there are two types of sublattice A and B. The tight binding Hamiltonian for graphene-based systems is written in terms of creation and annihilation of electrons in these A and B sublattice. Suppose our twisted bilayer graphene resulted from conventional bilayer graphene when the lower layer was twisted by angle $\theta_{1}=-0.5\theta_c$ along with upper layer being twisted by angle $\theta_{2}=0.5\theta_c$. A simple tight binding Hamiltonian for such TBG, including  nearest neighbour hopping, next nearest neighbour hopping along with on-site Coulomb interaction in  intra-layer contribution and environment dependent interlayer hopping in inter-layer contribution,   will have the form as given by equations \ref{eq:1a}, \ref{eq:1b}, \ref{eq:1c}.

\numparts\begin{eqnarray}
		H^{TBG} = H^{intra} + H^{inter} \label{eq:1a}\qquad\qquad\\
		\nonumber\\
		H^{intra}=-\sum_{\langle i,j\rangle ,s,l}t_1(a^{\dagger}_{i,s,l}b_{j,s,l}+ h.c.)\qquad\qquad\nonumber\\
		-\sum_{\langle\langle i,j\rangle\rangle,s,l}t_2(a^{\dagger}_{i,s,l}a_{j,s,l}
		+b^{\dagger}_{i,s,l}b_{j,s,l}+h.c.)\qquad\nonumber\\
		+U\sum_{i,s,l}(a^{\dagger}_{i,s,l}a^{\dagger}_{i,-s,l}a_{i,-s,l}a_{i,s,l})\qquad\qquad\nonumber\\
		+U\sum_{i,s,l}(b^{\dagger}_{i,s,l}b^{\dagger}_{i,-s,l}b_{i,-s,l}b_{i,s,l}) \label{eq:1b}\qquad\qquad\\
		H^{inter}=-\sum_{i,j,s}t_{\perp}(\bi{R}^A_{i,1},\bi{R}^A_{j,2})(a^{\dagger}_{j,s,2}a_{i,s,1}+h.c.)\quad\quad\nonumber\\
		-\sum_{i,j,s}t_{\perp}(\bi{R}^B_{i,1},\bi{R}^B_{j,2})(b^{\dagger}_{j,s,2}b_{i,s,1}+h.c.)\quad\quad\nonumber\\
		-\sum_{i,j,s}t_{\perp}(\bi{R}^A_{i,1},\bi{R}^B_{j,2})(b^{\dagger}_{j,s,2}a_{i,s,1}+h.c.)\quad\quad\nonumber\\
		-\sum_{i,j,s}t_{\perp}(\bi{R}^B_{i,1},\bi{R}^A_{j,2})(a^{\dagger}_{j,s,2}b_{i,s,1}+h.c.) \label{eq:1c}\quad\quad		
	\end{eqnarray}\endnumparts

Here $i,j$ are index for position of lattice sites in a layer. $\langle i,j \rangle$ means that $i^{th}$ site and $j^{th}$  site can only be nearest neighbours of same layer. $\langle\langle i,j \rangle\rangle$ means that $i^{th}$ site and $j^{th}$  site can only be second nearest neighbours of same layer. $s$  is index for spin which can take only two values; $\uparrow$ and $\downarrow$. $l$  is the index which denotes the layer; $l=1$ for lower layer and $l=2$ for upper layer. $a^{\dagger}_{i,s,l}$ ($a_{i,s,l}$) creates (annihilates) an electron on sub-lattice A in layer with index $l$ on $i^{th}$ site with spin $s$. $b^{\dagger}_{i,s,l}$ ($b_{i,s,l}$) creates (annihilates) an electron on sub-lattice B in layer with index $l$ on $i^{th}$ site with spin $s$. $t_1$ is the intra-layer nearest neighbour hopping energy and it is taken to be same on all sites in both layers. $t_2$ is the intra-layer second nearest neighbour hopping energy and it is also taken to be same on all sites in both layers. $U$ is on-site Coulomb interaction energy which is due to Coulomb interaction between two electrons of opposite spin residing on same site and it is taken to be same on all sites in both layers. First term in $H^{intra}$ represents hopping energy of electrons due to hopping between nearest neighbours of same layer. Second term in $H^{intra}$ represents hopping energy of electrons due to hopping between second nearest neighbours of same layer. Third and Fourth terms in $H^{intra}$ represent the potential energy of electrons due to Coulomb interaction occurring between two electrons residing on the same site.  $\bi{R}^{(A)}_{i,l}$ or $\bi{R}^{(B)}_{i,l}$ denote the position of $i^{th}$ site in sub lattice A or B of layer with index $l$. $t_{\perp}(\bi{R}^{(A/B)}_{(i,1)},\bi{R}^{(A/B)}_{(j,2)})$ represent the interlayer hopping energy of an electron hopping between sites $\bi{R}^{(A/B)}_{(i,1)}$ of layer 1 and $\bi{R}^{(A/B)}_{(j,2)}$ of layer 2. First, second, third and fourth terms in $H^{inter}$ represents hopping energy of electrons due to interlayer hopping. $h.c.$ is for Hermitian conjugate of corresponding term contained in parentheses. Hamiltonian model given by equations \ref{eq:1a}, \ref{eq:1b}, \ref{eq:1c} is written in terms of real space positions of electrons and to obtain energy dispersion from this Hamiltonian it needs to be transformed from real space to wave-vector space. Hamiltonian model given by equations \ref{eq:2a}, \ref{eq:2b} is wave-vector space representation of Hamiltonian model given by equations \ref{eq:1a}, \ref{eq:1b}, \ref{eq:1c}, which is obtained after applying Fourier transformation.

\numparts\begin{eqnarray}
	H^{TBG}=\sum_{\bi{k},s}\left( \left(
	\begin{array}{cccc}
		a^{\dagger}_{\bi{k},s,1}	& b^{\dagger}_{\bi{k},s,1} & a^{\dagger}_{\bi{k},s,2}	& b^{\dagger}_{\bi{k},s,2} 
	\end{array}
	\right) 
	H^{TBG}_{\bi{k},s}
	\left(
	\begin{array}{c}
		a_{\bi{k},s,1} \\	 b_{\bi{k},s,1} \\ a_{\bi{k},s,2}	\\ b_{\bi{k},s,2} 
	\end{array}
	\right)	\right) \label{eq:2a}\\
\fl	H^{TBG}_{\bi{k},s}=\left(
	\begin{array}{cccc}
		-t_2f_{2,\bi{k},1}+U\langle n^a_{-s,1}\rangle & -t_1f_{1,\bi{k},1} & -t^{AA*}_{\perp,\bi{k}} & -t^{AB*}_{\perp,\bi{k}}\\
		-t_1f^{*}_{1,\bi{k},1} & -t_2f_{2,\bi{k},1}+U\langle n^b_{-s,1}\rangle & -t^{BA*}_{\perp,\bi{k}} & -t^{BB*}_{\perp,\bi{k}}\\
		-t^{AA}_{\perp,\bi{k}} & -t^{BA}_{\perp,\bi{k}} & -t_2f_{2,\bi{k},2}+U\langle n^a_{-s,2}\rangle & -t_1f_{1,\bi{k},2}\\
		-t^{AB}_{\perp,\bi{k}} & -t^{BB}_{\perp,\bi{k}} & -t_1f^{*}_{1,\bi{k},2} & -t_2f_{2,\bi{k},2}+U\langle n^b_{-s,2}\rangle
	\end{array}
	\right)  \label{eq:2b}\qquad	 
\end{eqnarray}\endnumparts

Various terms involved in equations \ref{eq:2a}, \ref{eq:2b} are explained as follows. $\bi{k}$ is wave vector. $a^{\dagger}_{\bi{k},s,l}$ $(a_{\bi{k},s,l})$  creates (annihilates) an electron with wave-vector $\bi{k}$ and spin $s$ on sublattice-A of layer denoted by index-$l$; $b^{\dagger}_{\bi{k},s,l}$ $ (b_{\bi{k},s,l})$  creates (annihilates) an electron with wave-vector $\bi{k}$ and spin $s$ on sublattice-B of layer denoted by index-$l$. $t_1f_{1,\bi{k},l}$ is contribution from nearest neighbour intra-layer hopping in layer denoted by index $l$. $t_2f_{2,\bi{k},l}$ is contribution from second nearest neighbour intra-layer hopping in layer denoted by index $l$. $f_{1,\bi{k},l}$ is characteristic function for nearest neighbour intra-layer hopping in graphene layer denoted by index $l$. $f_{2,\bi{k},l}$ is characteristic function for second nearest neighbour intra-layer hopping in graphene layer denoted by index $l$. $U\langle n^a_{-s,l}\rangle$ is mean-field approximated contribution from onsite coulomb interaction occuring on sublattice-A of layer denoted by index-$l$. $U\langle n^b_{-s,l}\rangle$ is mean-field approximated contribution from onsite coulomb interaction occuring on sublattice-B of layer denoted by index-$l$. $\langle n^a_{-s,l}\rangle$ is average density per site of electrons of spin $-s$  on sublattice-A of layer denoted by index-$l$. $\langle n^b_{-s,l}\rangle$ is average density per site of electrons of spin $-s$  on sublattice-B of layer denoted by index-$l$. $t^{AA}_{\perp}(\bi{k})$ is contribution from inter-layer hopping taking place between carbon atoms of A-sublattice in lower layer and A-sublattice in upper layer. $t^{AB}_{\perp}(\bi{k})$ is contribution from inter-layer hopping taking place between carbon atoms of A-sublattice in lower layer and B-sublattice in upper layer. $t^{BA}_{\perp}(\bi{k})$ is contribution from inter-layer hopping taking place between carbon atoms of B-sublattice in lower layer and A-sublattice in upper layer. $t^{BB}_{\perp}(\bi{k})$ is contribution from inter-layer hopping taking place between carbon atoms of B-sublattice in lower layer and B-sublattice in upper layer. $*$ sign on any component means complex conjugate of that component. The functional form of various abbreviated terms involved in $H^{TBG}_{\bi{k},s}$ is given in equations:\ref{eq:3}, \ref{eq:4},  \ref{eq:5}, \ref{eq:6}, \ref{eq:7}, \ref{eq:8},  \ref{eq:9}, \ref{eq:10}, \ref{eq:11}, \ref{eq:12}. 

\begin{eqnarray}
	\bi{k}=k_x\hat{x}+k_y\hat{y} \label{eq:3}\qquad\qquad\qquad\\
	k^{r,l}_x=k_x\cos{\theta_{l}} +k_y\sin{\theta_{l}} \label{eq:4}\qquad\qquad\qquad\\
	k^{r,l}_y=-k_x\sin{\theta_{l}} +k_y\cos{\theta_{l}} \label{eq:5}\qquad\qquad\qquad\\
	f_{1,\bi{k},l}=\;\sum_{n=1}^{n=3}\rme^{\rm{i}\bi{k}\cdot\bdelta_{n,1,A,l}}=\;  \rme^{\frac{-\rm{i}a_o}{2}\left(\frac{k^{r,l}_x}{\sqrt{3}}\right)}\left\{ 2\cos\left(\frac{a_o k^{r,l}_y}{2}\right)+\rme^{\frac{\rm{i}a_o}{2}\left(\sqrt{3}k^{r,l}_x\right)}\right\} \label{eq:6}\qquad\\
	f_{2,\bi{k},l}=\;\sum_{n=1}^{n=6}\rme^{\rm{i}\bi{k}\cdot\bdelta_{n,2,A,l}}=\; \left\{ 4\cos\left(\frac{a_o k^{r,l}_y}{2}\right).\cos\left(\frac{\sqrt{3}a_o k^{r,l}_x}{2}\right)+2\cos\left(a_o k^{r,l}_y\right)\right\} \label{eq:7}\qquad\\ 
\fl	t_{\perp}(\bi{R}_{i,1},\bi{R}_{j,2})=\beta_{1}\exp{\left(-\beta_{2}\left(\frac{\left|\bi{R}^{\parallel}_{i,1}-\bi{R}^{\parallel}_{j,2} \right|}{\beta_{3}}\right)^{\beta_{4}} \right)}\times \exp{\left(-\beta_{5}\left(\frac{\left|z_{i,1}-z_{j,2} \right|-d_{AB}}{\beta_{6}}\right)^{\beta_{7}} \right)} \label{eq:8}\qquad\\
	t^{AA}_{\perp}(\bi{k})=\frac{1}{N_S}\sum^{|\bi{R}^{A\parallel}_{i,1}-\bi{R}^{A\parallel}_{j,2}|\leq R^{\parallel}_0}_{\langle\bi{R}^A_{i,1}\rangle_S,\bi{R}^A_{j,2}}t_{\perp}\left(\bi{R}^A_{i,1},\bi{R}^A_{j,2}\right)\rme^{\rm{i}\bi{k}\cdot\left(\bi{R}^{A\parallel}_{i,1}-\bi{R}^{A\parallel}_{j,2}\right)} \label{eq:9}\qquad\\
	t^{AB}_{\perp}(\bi{k})=\frac{1}{N_S}\sum^{|\bi{R}^{A\parallel}_{i,1}-\bi{R}^{B\parallel}_{j,2}|\leq R^{\parallel}_0}_{\langle\bi{R}^A_{i,1}\rangle_S,\bi{R}^B_{j,2}}t_{\perp}\left(\bi{R}^A_{i,1},\bi{R}^B_{j,2}\right)\rme^{\rm{i}\bi{k}\cdot\left(\bi{R}^{A\parallel}_{i,1}-\bi{R}^{B\parallel}_{j,2}\right)} \label{eq:10}\qquad\\
	t^{BA}_{\perp}(\bi{k})=\frac{1}{N_S}\sum^{|\bi{R}^{B\parallel}_{i,1}-\bi{R}^{A\parallel}_{j,2}|\leq R^{\parallel}_0}_{\langle\bi{R}^B_{i,1}\rangle_S,\bi{R}^A_{j,2}}t_{\perp}\left(\bi{R}^B_{i,1},\bi{R}^A_{j,2}\right)\rme^{\rm{i}\bi{k}\cdot\left(\bi{R}^{B\parallel}_{i,1}-\bi{R}^{A\parallel}_{j,2}\right)} \label{eq:11}\qquad\\
    t^{BB}_{\perp}(\bi{k})=\frac{1}{N_S}\sum^{|\bi{R}^{B\parallel}_{i,1}-\bi{R}^{B\parallel}_{j,2}|\leq R^{\parallel}_0}_{\langle\bi{R}^B_{i,1}\rangle_S,\bi{R}^B_{j,2}}t_{\perp}\left(\bi{R}^B_{i,1},\bi{R}^B_{j,2}\right)\rme^{\rm{i}\bi{k}\cdot\left(\bi{R}^{B\parallel}_{i,1}-\bi{R}^{B\parallel}_{j,2}\right)} \label{eq:12}\qquad 
\end{eqnarray}	

$k_x$ is x-component and $k_y$ is y-component of wave vector $\bi{k}$. $\bdelta_{n,1,A,l}$ represent a translation vector to nearest neighbour of A-sublattice site in layer denoted by index $l$, $n=1,2,3$ is index corresponding to three nearest neighbours in a graphene layer. $\bdelta_{n,2,A,l}$ represent a translation vector to second nearest neighbour of A-sublattice site in layer denoted by index $l$, $n=1,2,3,4,5,6$ is index corresponding to six second nearest neighbours in a graphene layer. $t_{\perp}(\bi{R}_{i,1},\bi{R}_{j,2})$  represent environment dependent interlayer hopping integral corresponding to interlayer hopping taking place between a carbon atom of lower layer  at planar position denoted by $\bi{R}_{i,1}$ and other carbon atom of upper layer at planar position denoted by $\bi{R}_{j,2}$. $\left|\bi{R}^{\parallel}_{i,1}-\bi{R}^{\parallel}_{j,2} \right|$ is planar component of distance between  a carbon atom of lower layer at position denoted by $\bi{R}_{i,1}$ and other carbon atom of upper layer at position denoted by $\bi{R}_{j,2}$. $\left|z_{i,1}-z_{j,2} \right|$ is z component of distance between a carbon atom of lower layer at position denoted by $\bi{R}_{i,1}$ and other carbon atom of upper layer at position denoted by $\bi{R}_{j,2}$. $d_{AB}$ is interlayer distance in AB-stacked bilayer graphene. $\beta_{1}$, $\beta_{2}$, $\beta_{3}$, $\beta_{4}$, $\beta_{5}$, $\beta_{6}$ and $\beta_{7}$ are parameters which are estimated by fitting calculated dispersion and density of states with experimentally observed dispersion and density of states in AB-stacked bilayer graphene and some TBG systems. Supercell of TBG corresponding to commensurate moire period equal to $L_c$ contains $N_S$ lattice points of each of $A_1$, $B_1$, $A_2$ and $B_2$ sublattices, here $N_S=L^2_c$ and $L_c$ is in units of $a_o$. $R^{\parallel}_0$ denote the maximum planer distance between two carbon atoms of TBG upto which the interlayer hopping occurs.$\sum^{|\bi{R}^{A\parallel}_{i,1}-\bi{R}^{A\parallel}_{j,2}|\leq R^{\parallel}_0}_{\langle\bi{R}^{A/B}_{i,1}\rangle_S,\bi{R}^{A/B}_{j,2}}$ means that sublattice site denoted by position $\bi{R}^{A/B}_{i,1}$ is restricted to single supercell of TBG and sublattice site denoted by $\bi{R}^{A/B}_{j,2}$ can be only those sites whose planar distance from sublattice site denoted by position $\bi{R}^{A/B}_{i,1}$ is less than or equal to $R^{\parallel}_0$. After computing various components of Hamiltonian $H^{TBG}_{\bi{k},s}$, the eigen values of the Hamiltonian $H^{TBG}_{\bi{k},s}$ can easily be computed which will create the quasi particle energy dispersion and from quasi particle energy dispersion the density of states can easily be computed.

\section{Hamiltonian Parameters} \label{sec4}
To estimate the values of parameters $t_1$, $t_2$, $U$, $R^{\parallel}_0$, $\beta_1$, $\beta_2$, $\beta_3$, $\beta_4$ and $\beta_5$, we tried to fit our computationally obtained quasi particle energy dispersion with experimentally observed dispersion for monolayer graphene, AB-stacked bilayer graphene and some TBG systems.\\
In same basis as chosen for TBG, the Hamiltonian for monolayer graphene will have the form as given by equation \ref{eq:13}.
\begin{equation} \label{eq:13}
	H^{mono}_{\bi{k},s }=\left(
	\begin{array}{cc}
		-t_2f_{2,\bi{k}}+U\langle n^a_{-s}\rangle & -t_1f_{1,\bi{k}}\\
		-t_1f^{*}_{1,\bi{k}} & -t_2f_{2,\bi{k}}+U\langle n^b_{-s}\rangle
	\end{array}
	\right) 
\end{equation}

\begin{figure}[htp]
 	\centering
 	\subfloat[]{\includegraphics[width=0.5\textwidth,height=2 in]{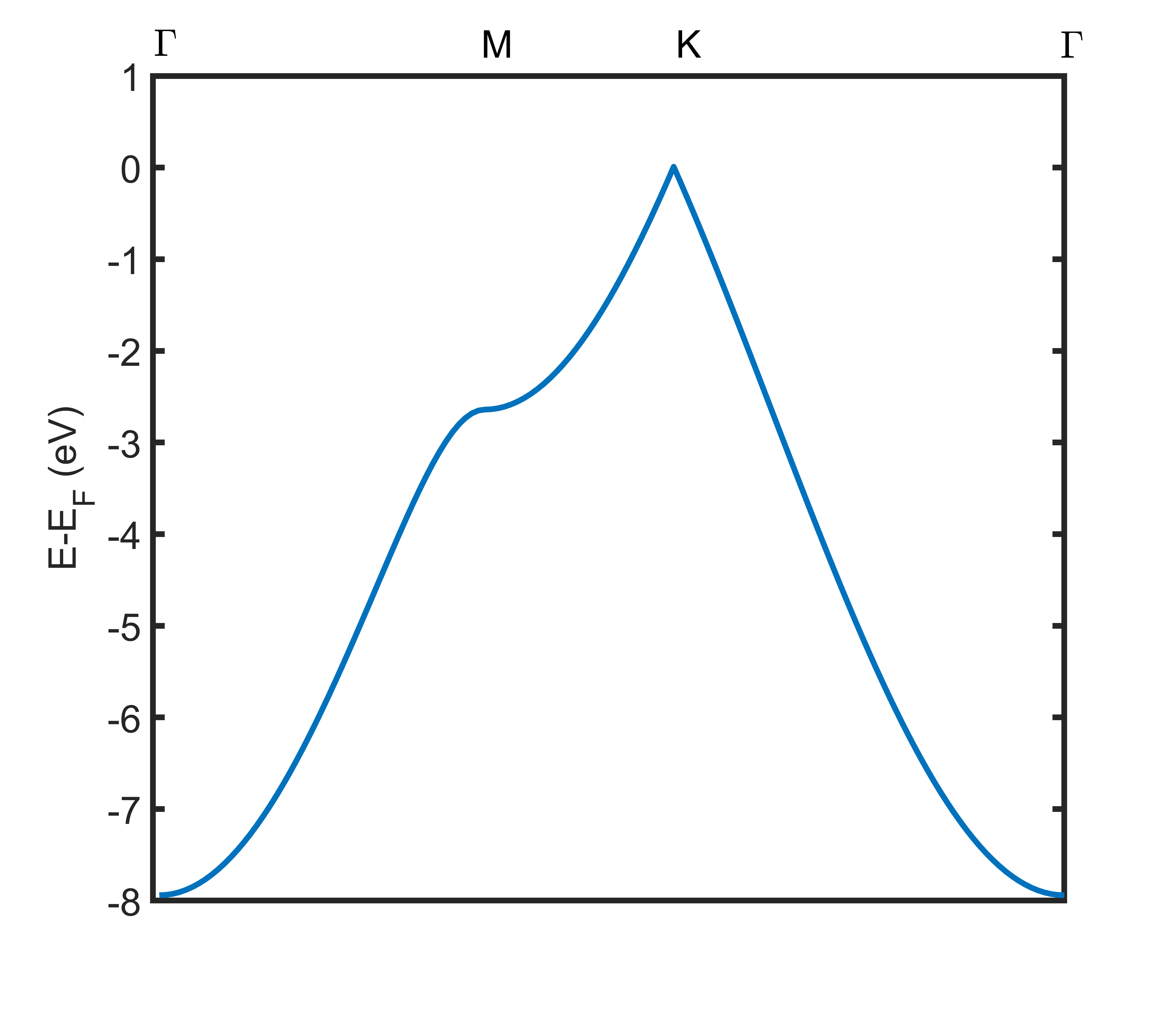}\label{monoG_dispersion1}}
 	\subfloat[]{\includegraphics[width=0.5\textwidth,height=2 in]{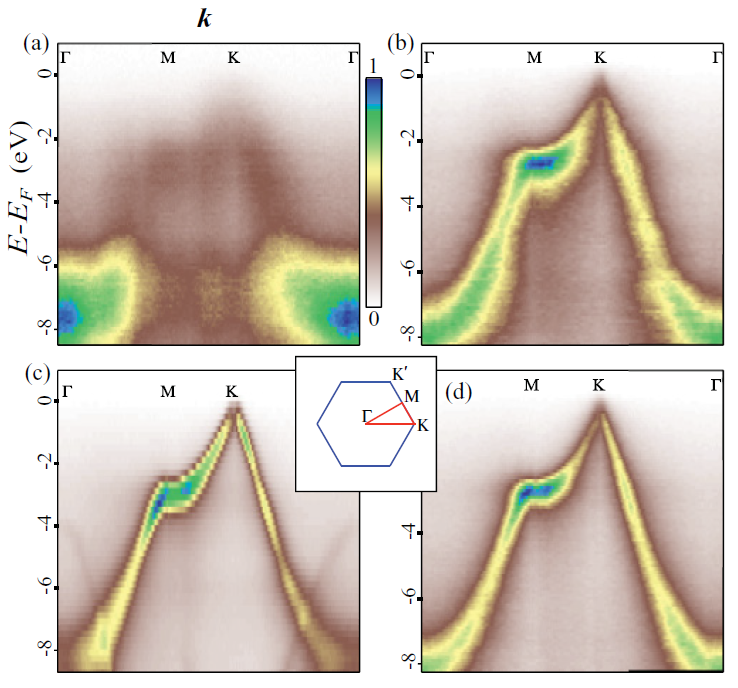}\label{monoG_dispersion2}}
 	\caption{ (a) Computed dispersion for monolayer graphene, (b) experimentally observed dispersion for monolayer graphene presented in Figure 2 of article of reference\cite{2011-Kevin}}		
 \end{figure}

 Here $f_{1,\bi{k}}$ and $f_{2,\bi{k}}$ are counterpart of  $f_{1,\bi{k},l}$ and $f_{2,\bi{k},l}$ respectively corresponding to untwisted graphene layer i.e., for $\theta_{l}=0$. $\langle n^a_{-s}\rangle=\langle n^b_{-s}\rangle=n_{in}\approx10^{-4}$ electron per atom. We computed eigen values of  $H^{mono}_{\bi{k},s }$ for various values of parameters $t_1$, $t_2$, and $U$. Computationally obtained quasi particle energy dispersion (shown in Figure:\ref{monoG_dispersion1}) fits best with experimentally observed dispersion(shown in Figure:\ref{monoG_dispersion2}) in monolayer graphene  for $t_1\approx2.65$ eV, $t_2\approx0$ eV, $U\approx9$ eV, which agrees well with previously reported values \cite{2011-Wehling}. 
 
\begin{figure}[htp]
	\centering
	\subfloat[]{\includegraphics[width=0.25\textwidth, height=3 in]{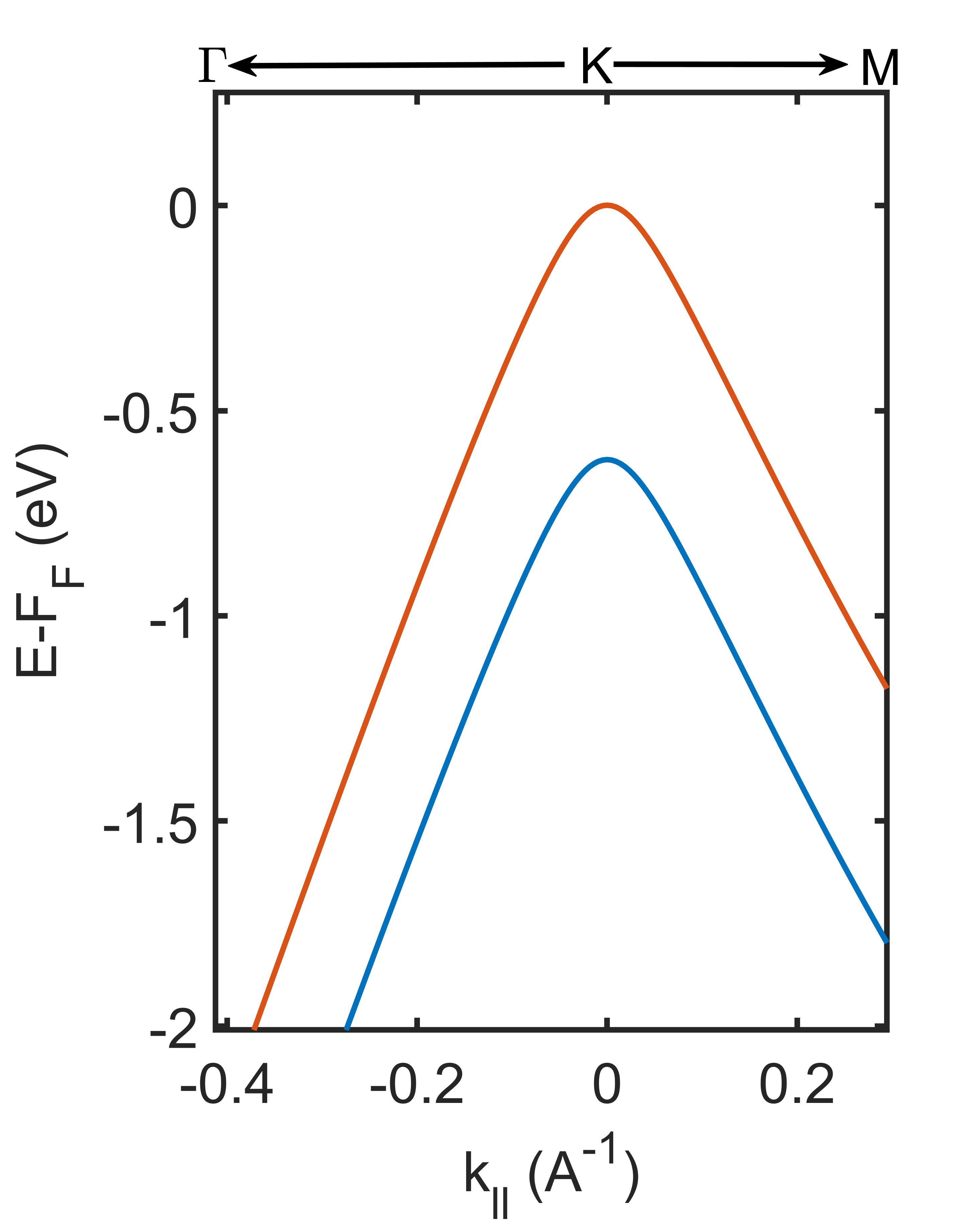}\label{ABG_dispersion1}}
	\subfloat[]{\includegraphics[width=0.25\textwidth, height=3 in]{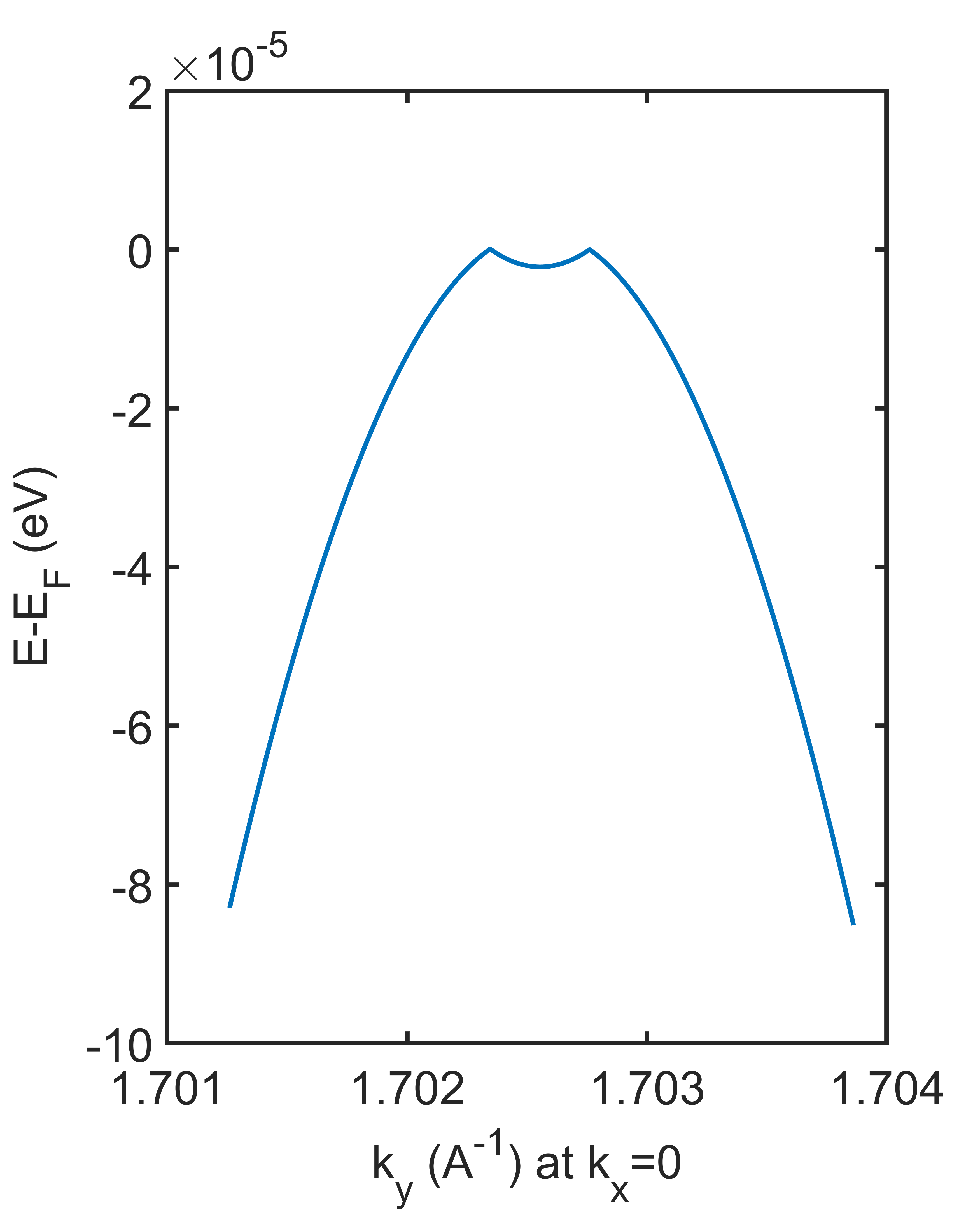}\label{ABG_dispersion2}}
	\subfloat[]{\includegraphics[width=0.5\textwidth, height=3 in]{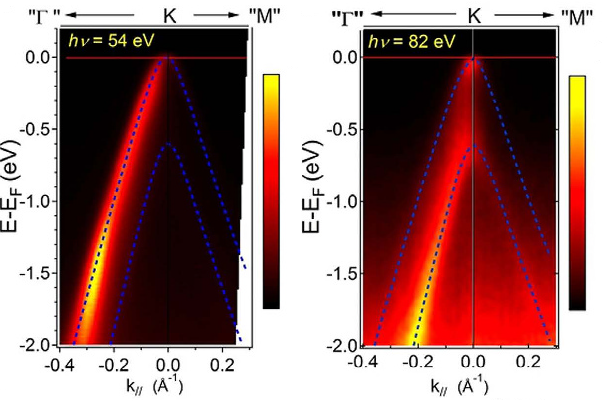}\label{ABG_dispersion3}}
	\caption{(a) Computed dispersion for AB-stacked bilayer graphene for $t_{\perp0}=0.62$ eV and $t_{\perp1}=0$ eV (b) Computed dispersion for AB-stacked bilayer graphene for $t_{\perp0}=0.62$ eV and $t_{\perp1}=0.01$ eV (c) experimentally observed ARPES dispersion for AB-stacked bilayer graphene presented in Figure 1(a), 1(c) of article of reference \cite{2015-Cheng}}		
\end{figure}

In same basis as chosen for TBG, the Hamiltonian for AB-stacked bilayer graphene will have the form as given by equation \ref{eq:14}.

\begin{equation}\label{eq:14}
\fl	H^{AB}_{\bi{k},s }=\left(
\begin{array}{cccc}
	-t_2f_{2,\bi{k}}+U\langle n^a_{-s}\rangle & -t_1f_{1,\bi{k}} & -t_{\perp1}f_{1,\bi{k}} & -t_{\perp0}\\
	-t_1f^{*}_{1,\bi{k}} & -t_2f_{2,\bi{k}}+U\langle n^b_{-s}\rangle & -t_{\perp1}f^{*}_{1,\bi{k}} & -t_{\perp1}f_{1,\bi{k}}\\
	-t_{\perp1}f^{*}_{1,\bi{k}} & -t_{\perp1}f_{1,\bi{k}} & -t_2f_{2,\bi{k}}+U\langle n^a_{-s}\rangle & -t_1f_{1,\bi{k}}\\
	-t_{\perp0} & -t_{\perp1}f^{*}_{1,\bi{k}} & -t_1f^{*}_{1,\bi{k}} & -t_2f_{2,\bi{k}}+U\langle n^b_{-s}\rangle
\end{array}
\right) 
\end{equation}
Here $t_{\perp0}=\beta_{1}=t_{\perp}(\bi{R}_{i,1},\bi{R}_{j,2})$ when $\left(\left|\bi{R}^{\parallel}_{i,1}-\bi{R}^{\parallel}_{j,2} \right|=0\,,\left|z_{i,1}-z_{j,2} \right|=d_{AB}\right)$; $t_{\perp1}=t_{\perp}(\bi{R}_{i,1},\bi{R}_{j,2})$ when $\left(\left|\bi{R}^{\parallel}_{i,1}-\bi{R}^{\parallel}_{j,2} \right|=a\,,\left|z_{i,1}-z_{j,2} \right|=d_{AB}\right)$. We used $t_1=2.65$ eV, $t_2=0$ eV and $U=9$ eV, and computed eigen values of  $H^{AB}_{\bi{k},s }$ for various values of parameters $t_{\perp0}$ and $t_{\perp1}$. Non zero values of $t_{\perp0}$ change the dispersion at Dirac point from linear to parabolic. Change in 
$t_{\perp0}$ changes the slope of dispersion at Dirac point and gap between two valence bands at Dirac point and this gap is almost equal to $t_{\perp0}$ (shown in Figure:\ref{ABG_dispersion1}, \ref{ABG_dispersion3}). Non zero values of $t_{\perp1}$ insert a concave deformation in dispersion at Dirac point (shown in Figure:\ref{ABG_dispersion2}). Computationally obtained quasi particle energy dispersion (shown in Figure:\ref{ABG_dispersion1}) fits best with experimentally observed dispersion (shown in Figure:\ref{ABG_dispersion3}) in AB-stacked bilayer graphene for $t_{\perp0}=0.62$ eV and $t_{\perp1}=0$ eV. 

$t_{\perp1}=0$ eV suggests that interlayer hopping $t_{\perp}(\bi{R}_{i,1},\bi{R}_{j,2})$ drops to zero when $\left|\bi{R}^{\parallel}_{i,1}-\bi{R}^{\parallel}_{j,2} \right|=a$, therefore $R^{\parallel}_0=a$. Since range of inter-layer hopping ($R^{\parallel}_0=a$) is small; after simulating the internal configurations of carbon atoms inside a supercell of TBG it becomes relatively easier to compute  $t^{AA}_{\perp}(\bi{k})$, $t^{AB}_{\perp}(\bi{k})$, $t^{BA}_{\perp}(\bi{k})$ and  $t^{BB}_{\perp}(\bi{k})$ for varying values of parameters $\beta_{2}$, $\beta_{3}$, $\beta_{4}$, $\beta_{5}$, $\beta_{6}$ and $\beta_{7}$. After computing the values of $t^{AA}_{\perp}(\bi{k})$, $t^{AB}_{\perp}(\bi{k})$, $t^{BA}_{\perp}(\bi{k})$ and  $t^{BB}_{\perp}(\bi{k})$, eigen values of Hamiltonian $H^{TBG}_{\bi{k},s }$ are computed very easily. From eigen values of $H^{TBG}_{\bi{k},s }$ dispersion and density of states are obtained. 
\begin{figure}[htp]
	\centering
	\subfloat[]{\includegraphics[width=0.5\textwidth, height=2.5 in]{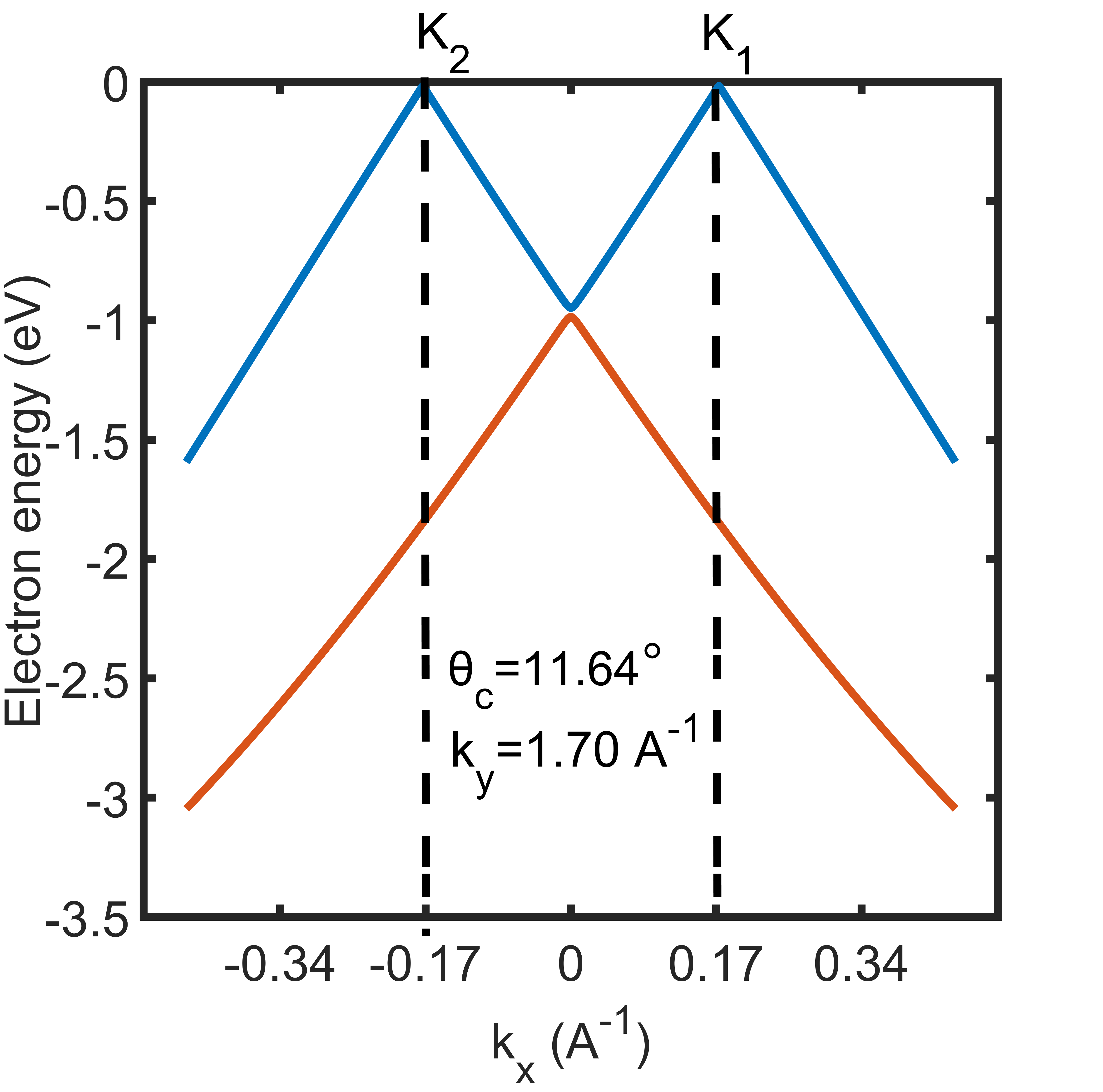}\label{TBG_dispersion1_thetaC=11.64_degree}}
	\subfloat[]{\includegraphics[width=0.5\textwidth, height=2.5 in]{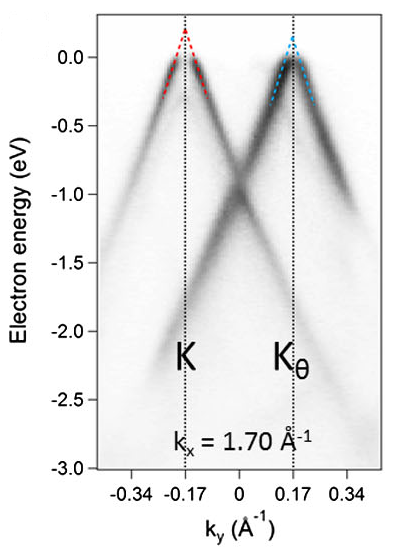}\label{TBG_dispersion2_thetaC=11.64_degree}}\\
	\subfloat[]{\includegraphics[width=0.5\textwidth, height=2.5 in]{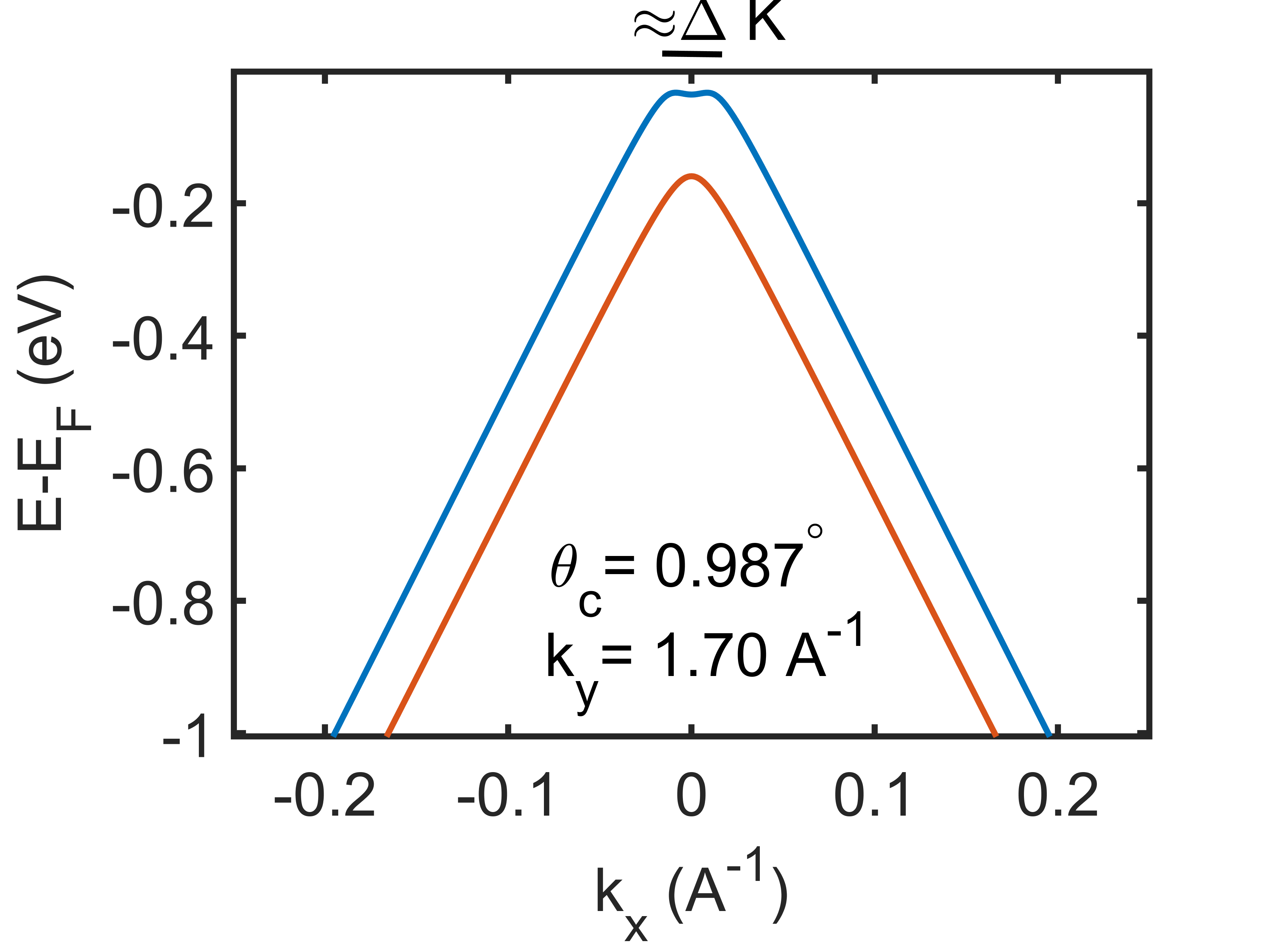}\label{TBG_dispersion_thetaC=0.98_degree}}
	\subfloat[]{\includegraphics[width=0.5\textwidth, height=2.5 in]{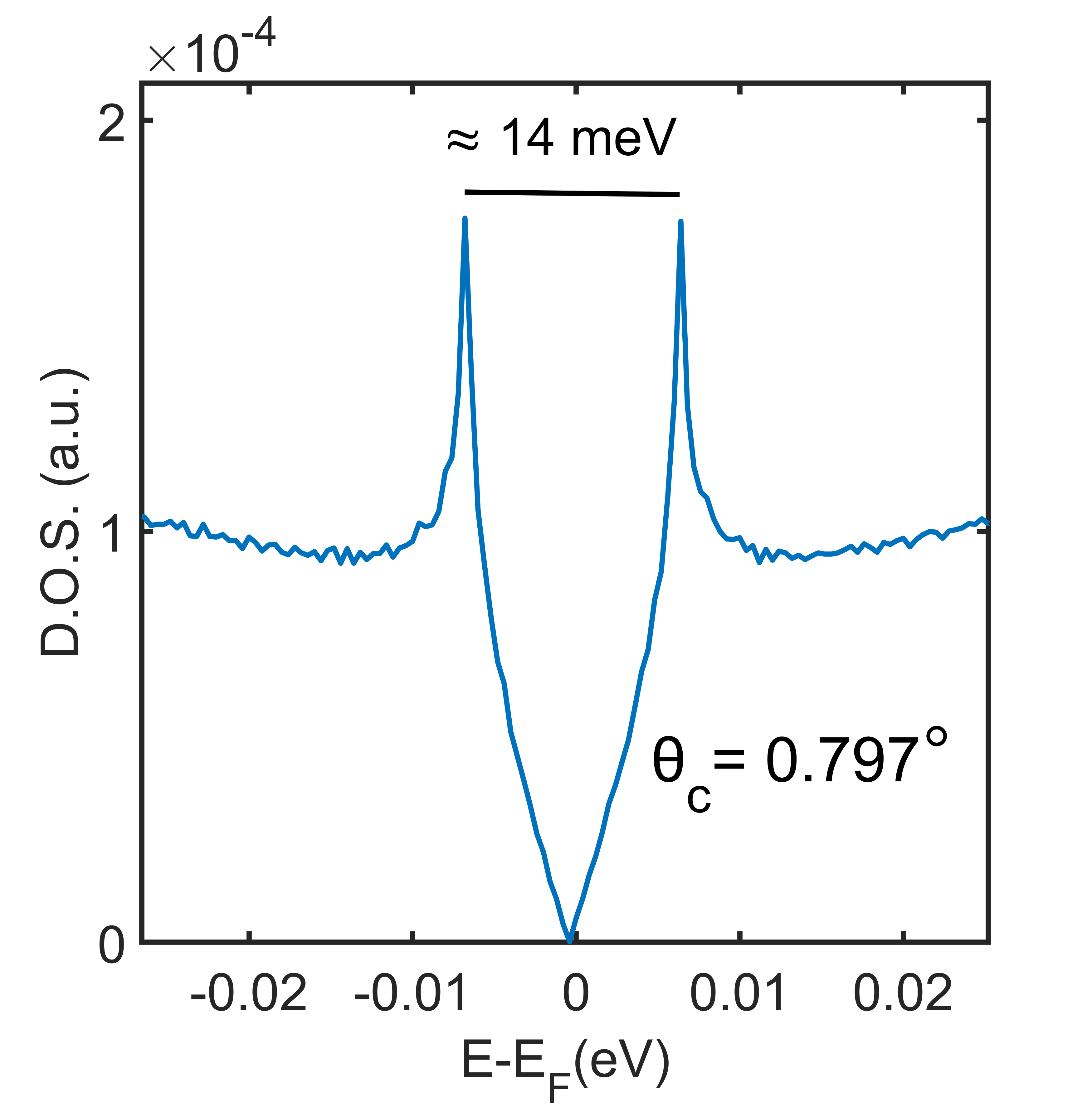}\label{TBG_DOS1_thetaC=0.79_degree}}
	\caption{(a) Computed dispersion for TBG with twist angle $11.64^{\circ}$ (b) Experimentally observed ARPES dispersion for TBG with twist angle $11.64^{\circ}$ \cite{2012-Taisuke}, (c) Computed dispersion for TBG with twist angle $0.98^{\circ}$ (in good agreement with experimentally observed dispersion presented in Figures:1(g), 2(c) of reference \cite{2020-Utama}) (d) Computed D.O.S. for TBG with twist angle $0.797^{\circ}$ (gap between two Van-Hove singularity peaks is in good agreement with experimentally observed  gap between two Van-Hove singularity peaks presented in Figure: 2(c) of reference \cite{2019-Alexander})}		
\end{figure}
To estimate the values of parameters $\beta_{2}$, $\beta_{3}$, $\beta_{4}$, $\beta_{5}$, $\beta_{6}$ and $\beta_{7}$ we tried to simultaneously fit; the computed dispersion with experimentally observed dispersion in TBG with twist angle $11.64^{\circ}$, the computed dispersion with experimentally observed dispersion in TBG with twist angle $0.98^{\circ}$ and  the gap between two Van-Hove singularity peaks in computed density of states and experimentally observed tunnelling conductance in TBG with twist angle $0.79^{\circ}$. Computationally obtained results fit best with experimentally observed results for $\beta_2=1.45$, $\beta_3=0.63$, $\beta_4=1.7$, $\beta_5=1.84$, $\beta_5=0.11$ and $\beta_5=1.5$.\\
Figure \ref{TBG_dispersion1_thetaC=11.64_degree} shows computed dispersion for TBG with twist angle $11.64^{\circ}$, which is in good agreement with experimentally observed dispersion shown in Figure \ref{TBG_dispersion2_thetaC=11.64_degree}. Figure \ref{TBG_dispersion2_thetaC=11.64_degree} show experimentally observed dispersion for TBG with twist angle $11.64^{\circ}$ presented in FIG. 1(c) of reference \cite{2012-Taisuke}. Figure \ref{TBG_dispersion_thetaC=0.98_degree} shows computed dispersion for TBG with twist angle $0.98^{\circ}$ which is in good agreement with  Fig:1(g), 2(c) of reference \cite{2020-Utama} showing experimentally observed dispersion for TBG with twist angle $0.98^{\circ}$. Figure \ref{TBG_DOS1_thetaC=0.79_degree} shows computed D.O.S. for TBG with twist angle $0.797^{\circ}$, the gap between two Van-Hove singularity peaks is in good agreement with  Fig:2(c) of reference \cite{2019-Alexander} showing experimentally observed gap between two Van-Hove singularity peaks for TBG with twist angle $0.79^{\circ}$. After estimation of all parameters we computed electronic energy dispersion and D.O.S. for many TBG systems, some of those results are presented in following section.

\section{Results and Discussion} \label{sec5}

\begin{figure} [htp]
	\centering
	\subfloat[]{\includegraphics[width=0.49\textwidth,height=1.8 in]{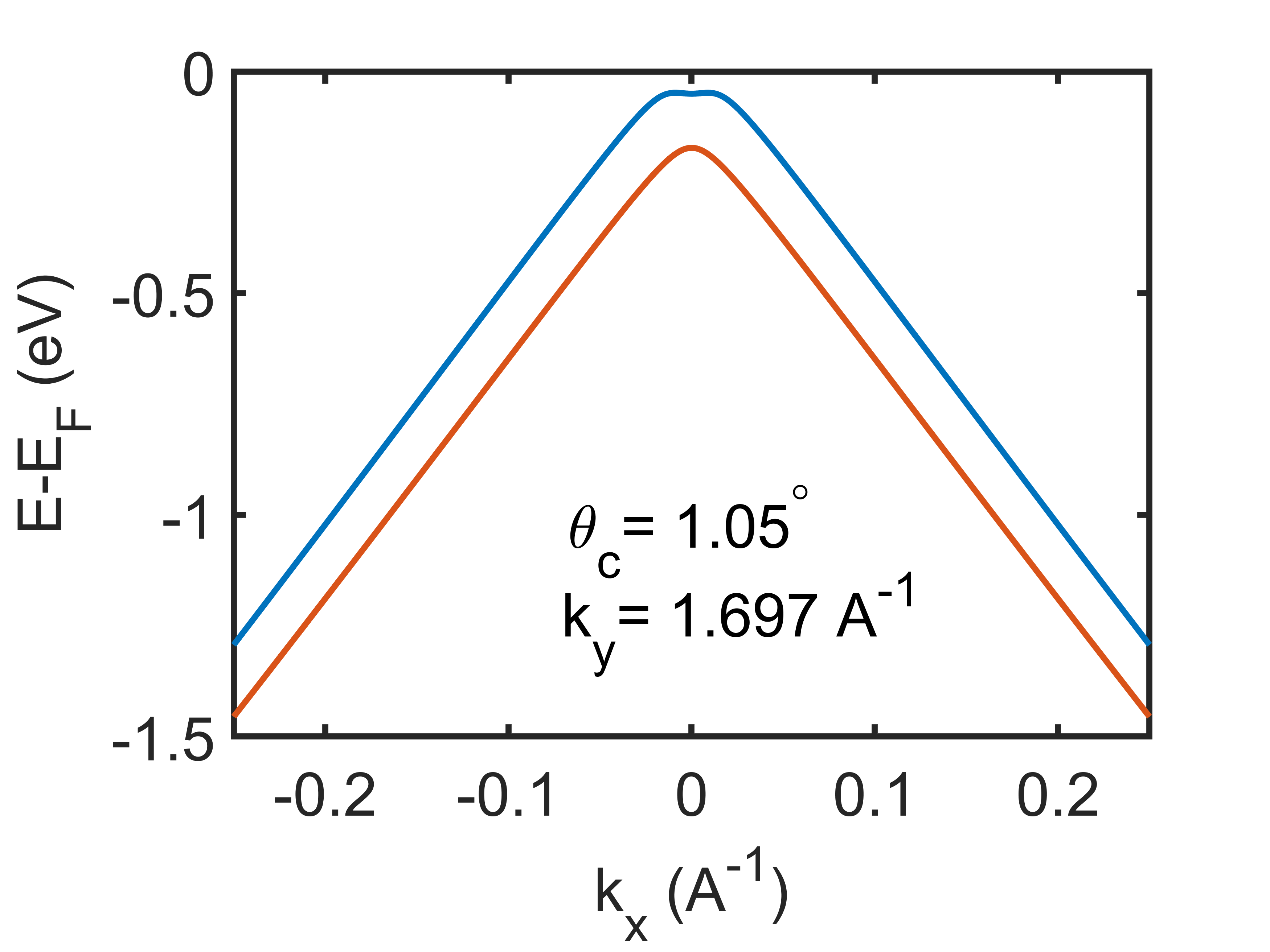}\label{TBG_dispersion_thetaC=1.05_degree}}
	\subfloat[]{\includegraphics[width=0.49\textwidth,height=1.8 in]{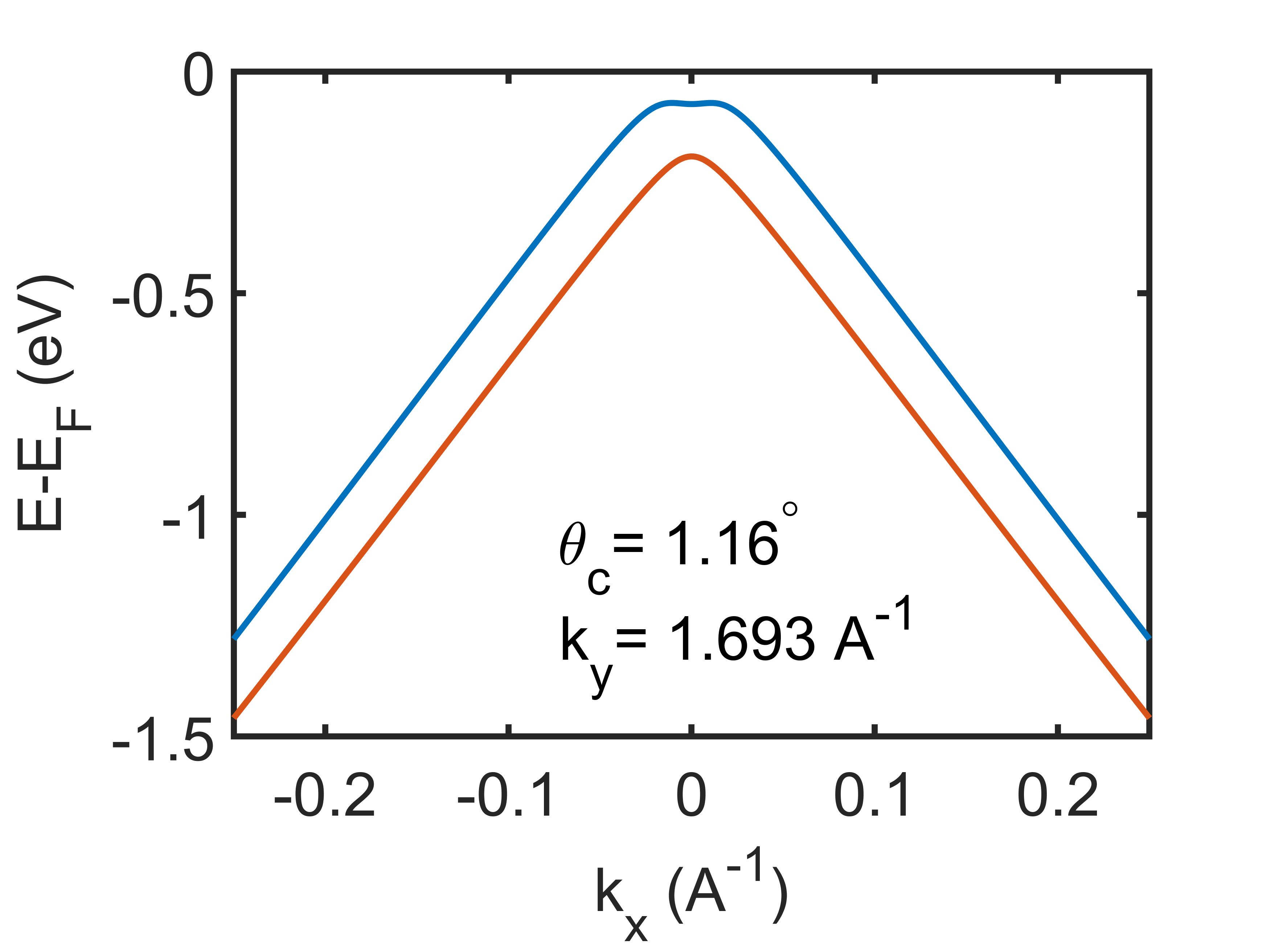}\label{TBG_dispersion_thetaC=1.16_degree}}\\
	\subfloat[]{\includegraphics[width=0.49\textwidth,height=1.8 in]{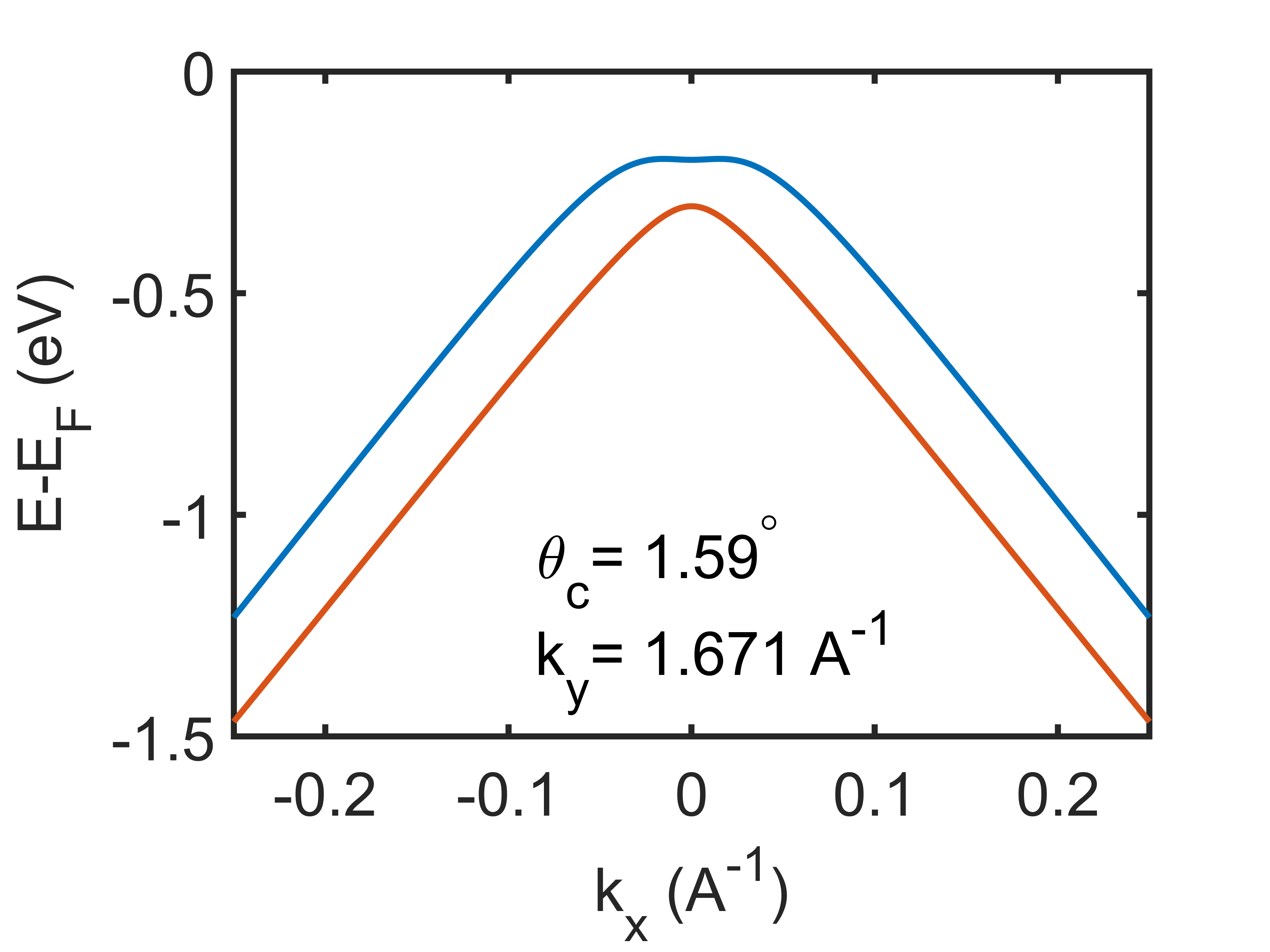}\label{TBG_dispersion_thetaC=1.59_degree}}
	\subfloat[]{\includegraphics[width=0.49\textwidth,height=1.8 in]{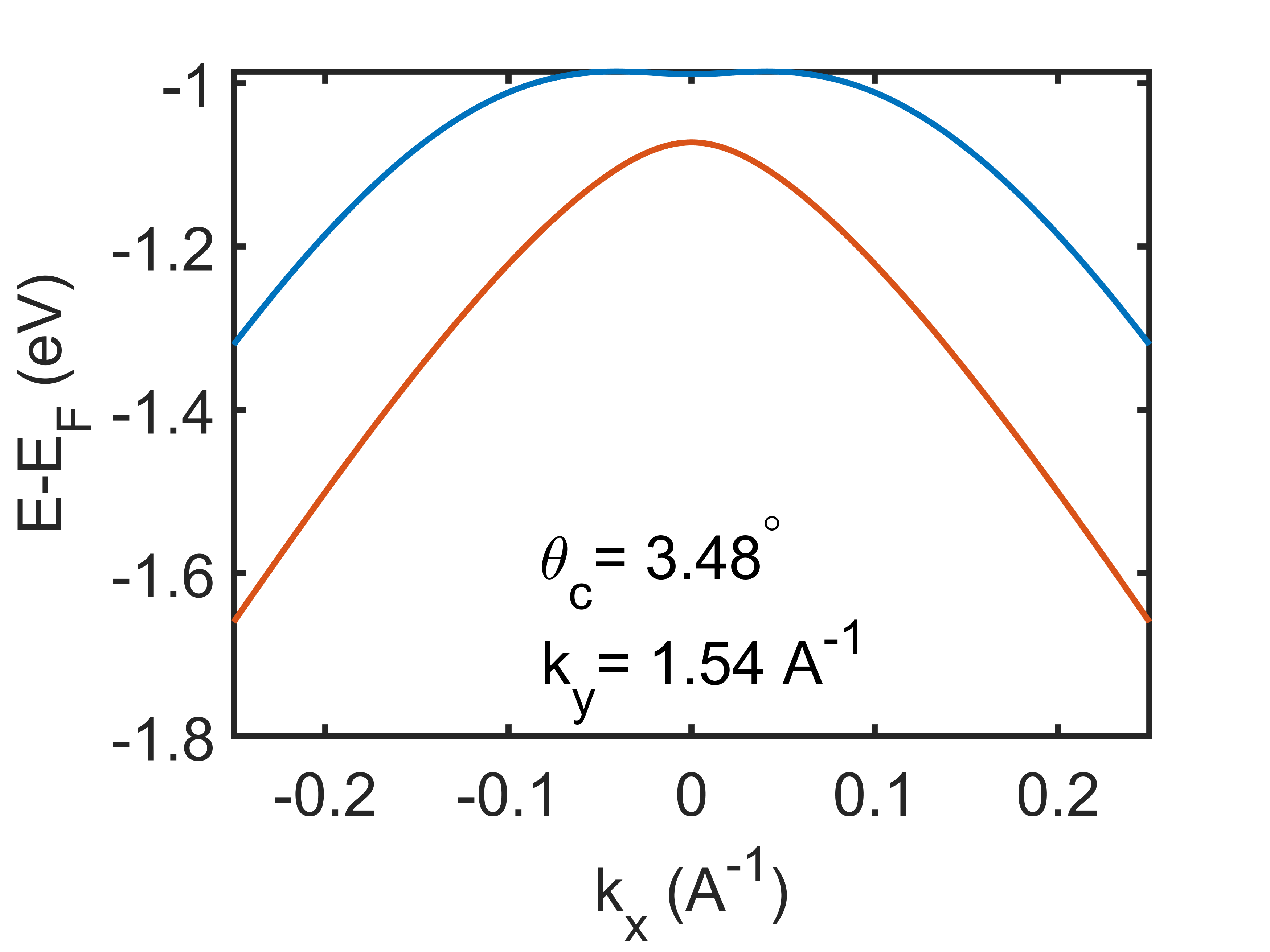}\label{TBG_dispersion_thetaC=3.48_degree}}
	\caption{Computed electronic energy dispersion for TBG with twist angle (a) $\theta_c=1.05^{\circ}$, (b) $\theta_c=1.16^{\circ}$, (c) $\theta_c=1.59^{\circ}$ and (d) $\theta_c=3.48^{\circ}$}\label{TBG_dispersion}		
\end{figure}		
Figures \ref{TBG_dispersion_thetaC=1.05_degree}, \ref{TBG_dispersion_thetaC=1.16_degree}, \ref{TBG_dispersion_thetaC=1.59_degree} and \ref{TBG_dispersion_thetaC=3.48_degree} show computed quasi particle energy dispersion for TBG with twist angle $\theta_c=1.05^{\circ}$, $\theta_c=1.16^{\circ}$, $\theta_c=1.59^{\circ}$ and $\theta_c=3.48^{\circ}$ respectively, all show presence of flat band. For TBG with twist $\theta_c=1.05^{\circ}$, flat band lies below $\approx 50 meV$ of Fermi level, for  TBG with twist $\theta_c=1.16^{\circ}$, flat band lies below $\approx 73 meV$ of Fermi level, for TBG with twist $\theta_c=1.59^{\circ}$, flat band lies below $\approx 198.5 meV$ of Fermi level and for TBG with twist $\theta_c=3.48^{\circ}$, flat band lies below $\approx 987 meV$ of Fermi level. As the twist angle increases, flat band move away from Fermi level. For TBG with twist $\theta_c=1.05^{\circ}$, flat band occurs near $k_y\approx 1.697 A^{-1}$ , for  TBG with twist $\theta_c=1.16^{\circ}$, flat band occurs near $k_y\approx 1.693 A^{-1}$, for TBG with twist $\theta_c=1.59^{\circ}$, flat band occurs near $k_y\approx 1.671 A^{-1}$ and for TBG with twist $\theta_c=3.48^{\circ}$, flat band occurs near $k_y\approx 1.54 A^{-1}$. The coordinates of chosen Dirac point are $\left(k_x=0 A^{-1},k_y= 1.7091 A^{-1}\right)$. As the twist angle increases, flat band move away from Dirac point. As the twist angle increases sharpness of flat band decreases and the dispersion moves towards curved structure. computationally obtained band dispersion reveal that flat band in TBG occurs very close to Dirac point of graphene and only along linear dimension of two dimensional wave vector space parallel to line connecting two closest Dirac points of two graphene layers of TBG.
\begin{figure}[htp]
	\centering
	\subfloat[]{\includegraphics[width=0.49\textwidth,height=1.8 in]{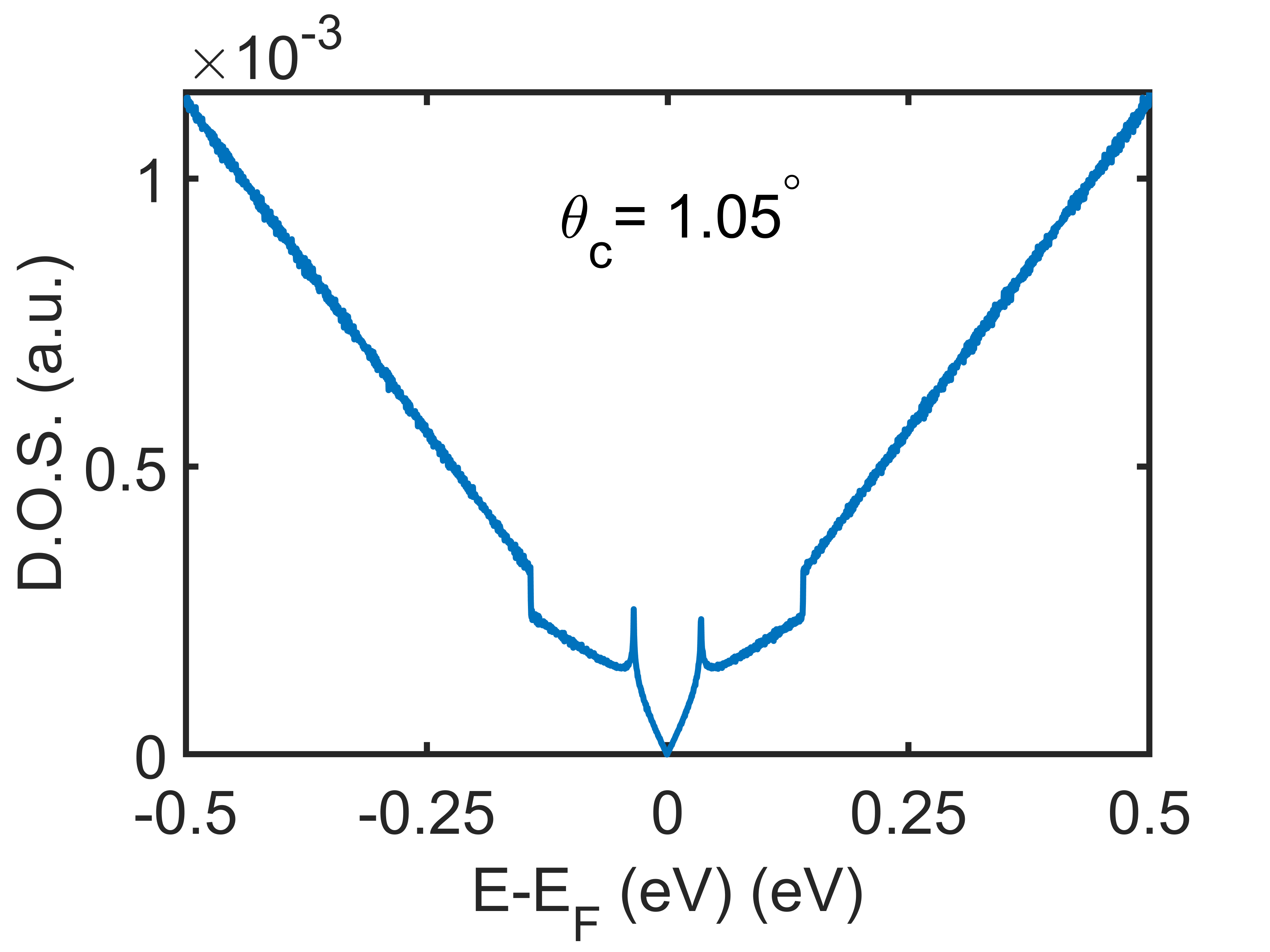}\label{TBG_DOS_thetaC=1.05_degree}}
	\subfloat[]{\includegraphics[width=0.49\textwidth,height=1.8 in]{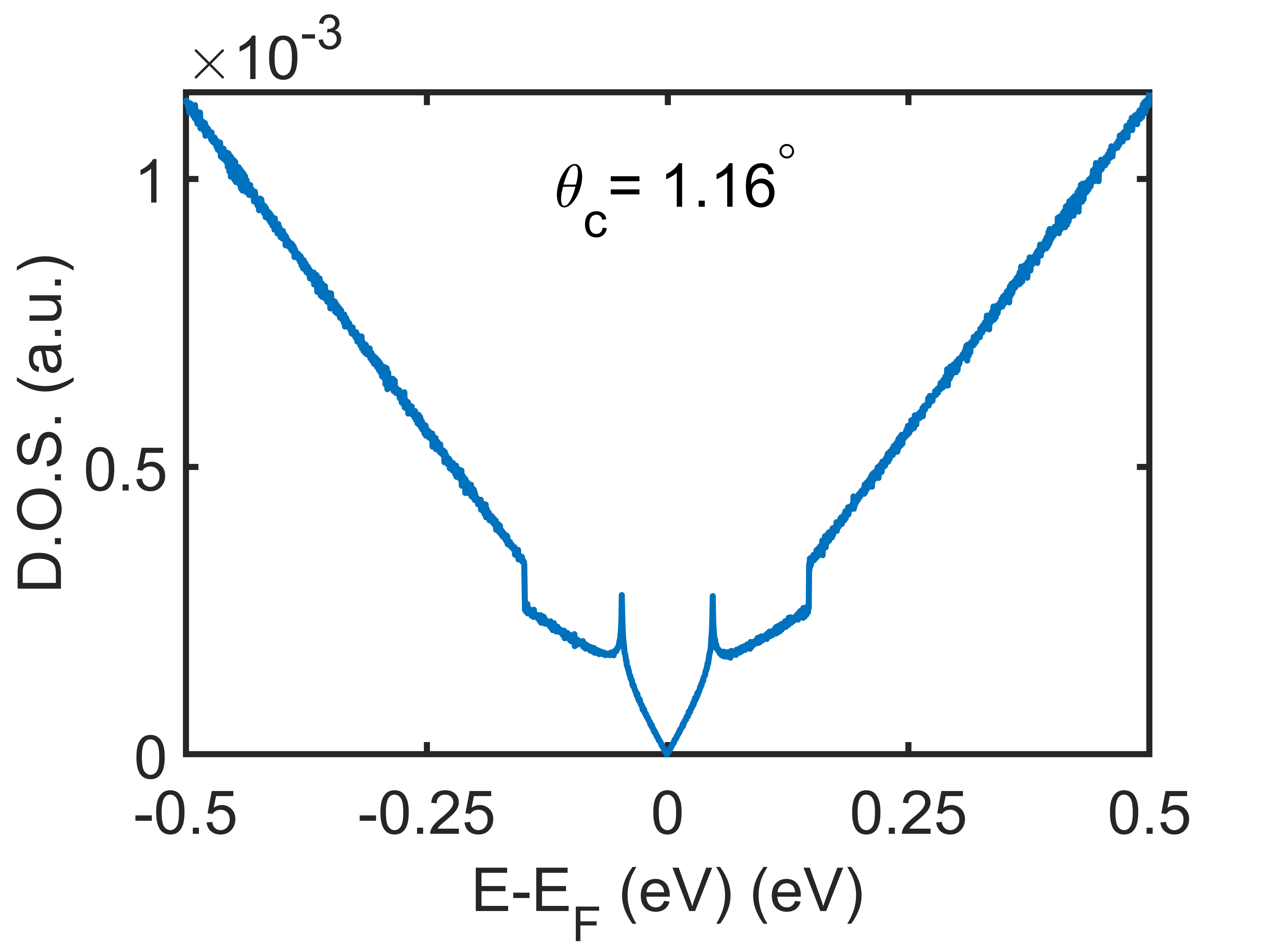}\label{TBG_DOS_thetaC=1.16_degree}}\\
	\subfloat[]{\includegraphics[width=0.49\textwidth,height=1.8 in]{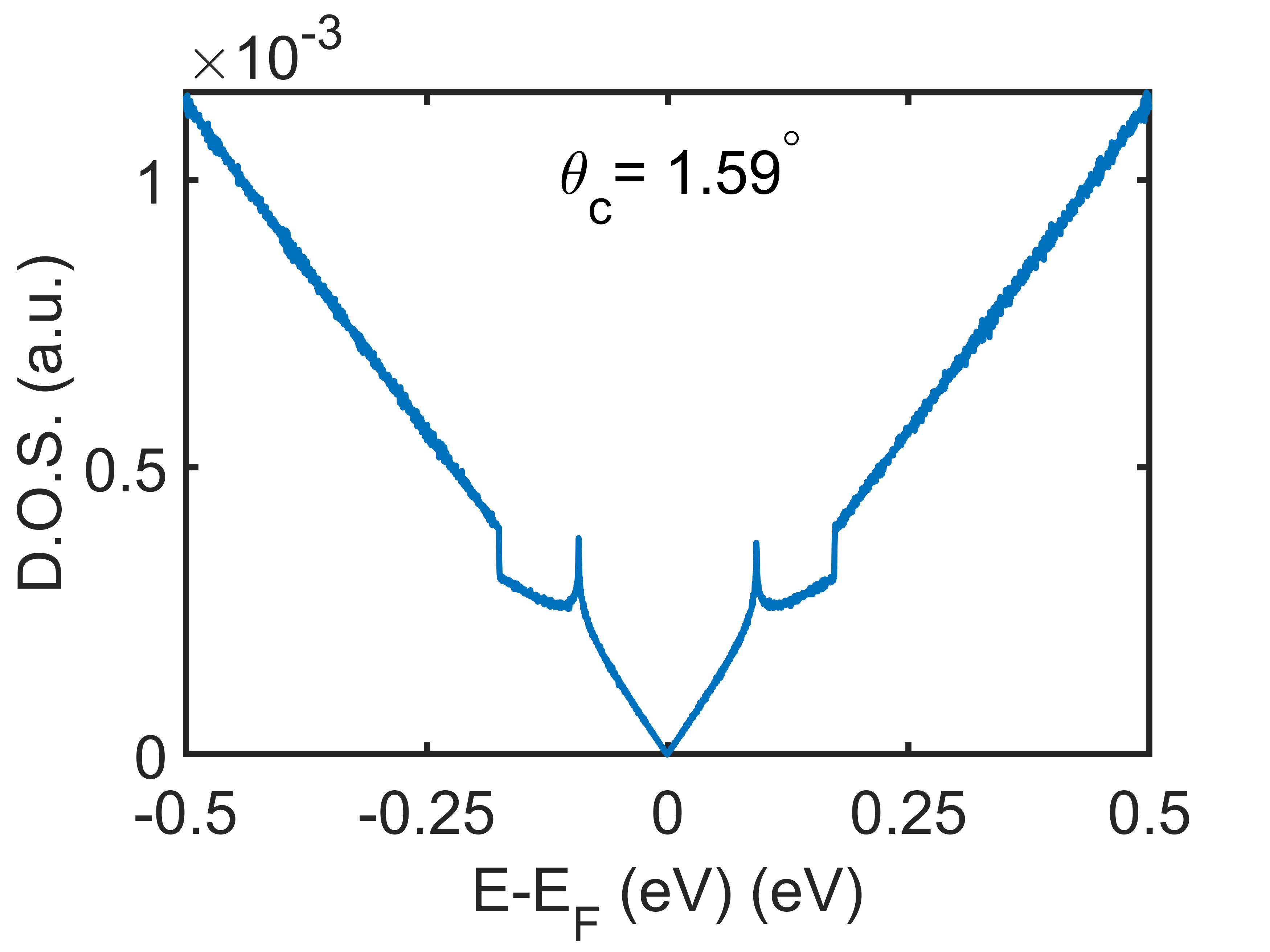}\label{TBG_DOS_thetaC=1.59_degree}}
	\subfloat[]{\includegraphics[width=0.49\textwidth,height=1.8 in]{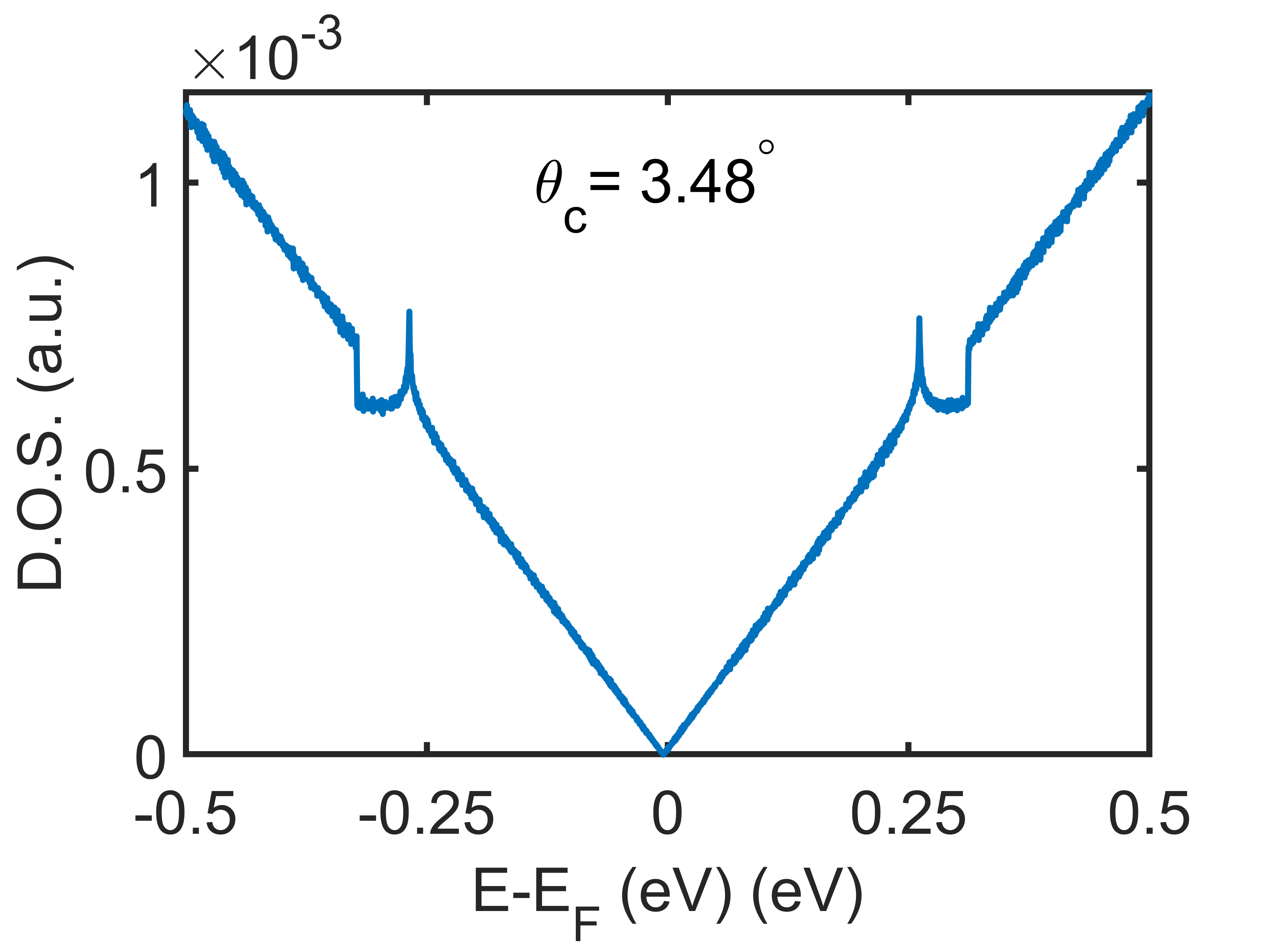}\label{TBG_DOS_thetaC=3.48_degree}}
	\caption{Computed electronic density of states for TBG with twist angle (a) $\theta_c=1.05^{\circ}$, (b) $\theta_c=1.16^{\circ}$, (c) $\theta_c=1.59^{\circ}$ and (d) $\theta_c=3.48^{\circ}$}\label{TBG_DOS}		
\end{figure}
Figures \ref{TBG_DOS_thetaC=1.05_degree}, \ref{TBG_DOS_thetaC=1.16_degree}, \ref{TBG_DOS_thetaC=1.59_degree} and \ref{TBG_DOS_thetaC=3.48_degree} show computed electronic density of states for TBG with twist angle $\theta_c=1.05^{\circ}$, $\theta_c=1.16^{\circ}$, $\theta_c=1.59^{\circ}$ and $\theta_c=3.48^{\circ}$ respectively, all show presence of two Van-Hove singularity peaks near Fermi level. As the twist angle increases, Van-Hove singularity peaks move away from Fermi level and gap between two Van-Hove singularity peaks increases. This behaviour in consistent with behaviour of flat band.  Computed dispersion and density of states for TBG show that instead of being restricted to TBG with discrete set of magic angles, flat band and Van-Hove singularities appear for all TBG systems having twist angle upto $\approx 3.5^{\circ}$ as observed experimentally. Even a simple tight binding Hamiltonian without electronic-correlations but with environment dependent interlayer hopping and incorporating complete internal configuration of carbon atoms inside a supercell can produce flat band dispersion and Van-Hove singularities near Fermi level for TBG. Flat band dispersion and Van-Hove singularities near Fermi gradually ruin as the twist angle increases. Shape of density of state near Fermi level indicate towards pseudo gap phase. Two Van-Hove singularities near Fermi level are consequence of  two flat bands near Dirac points, one above Fermi level and other below Fermi level.
\begin{figure}[htp]
	\centering
	\subfloat[]{\includegraphics[width=0.33\textwidth,height=2 in]{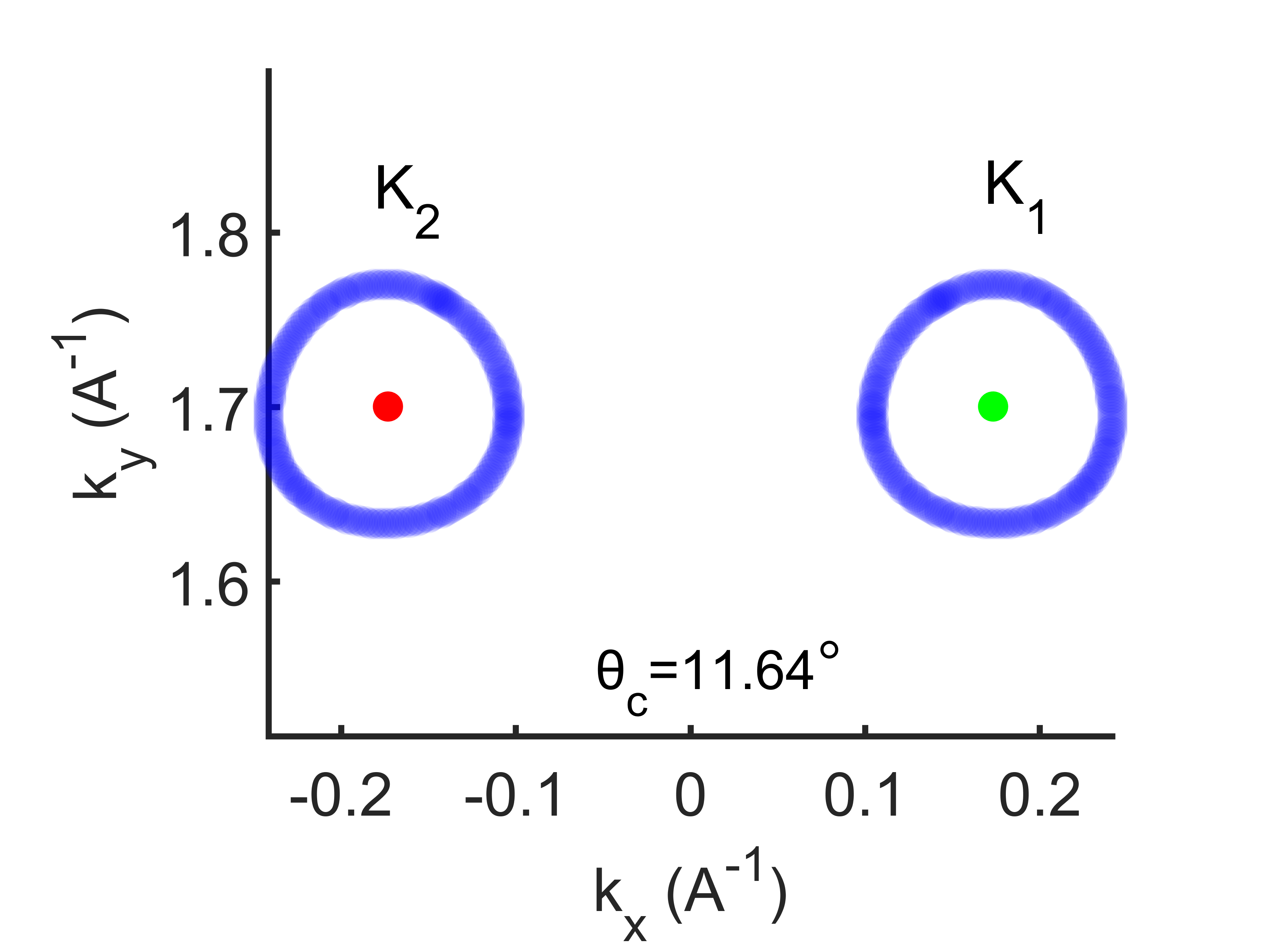}\label{Constant_energy_curve_thetaC=11.64_degree}}
	\subfloat[]{\includegraphics[width=0.33\textwidth,height=2 in]{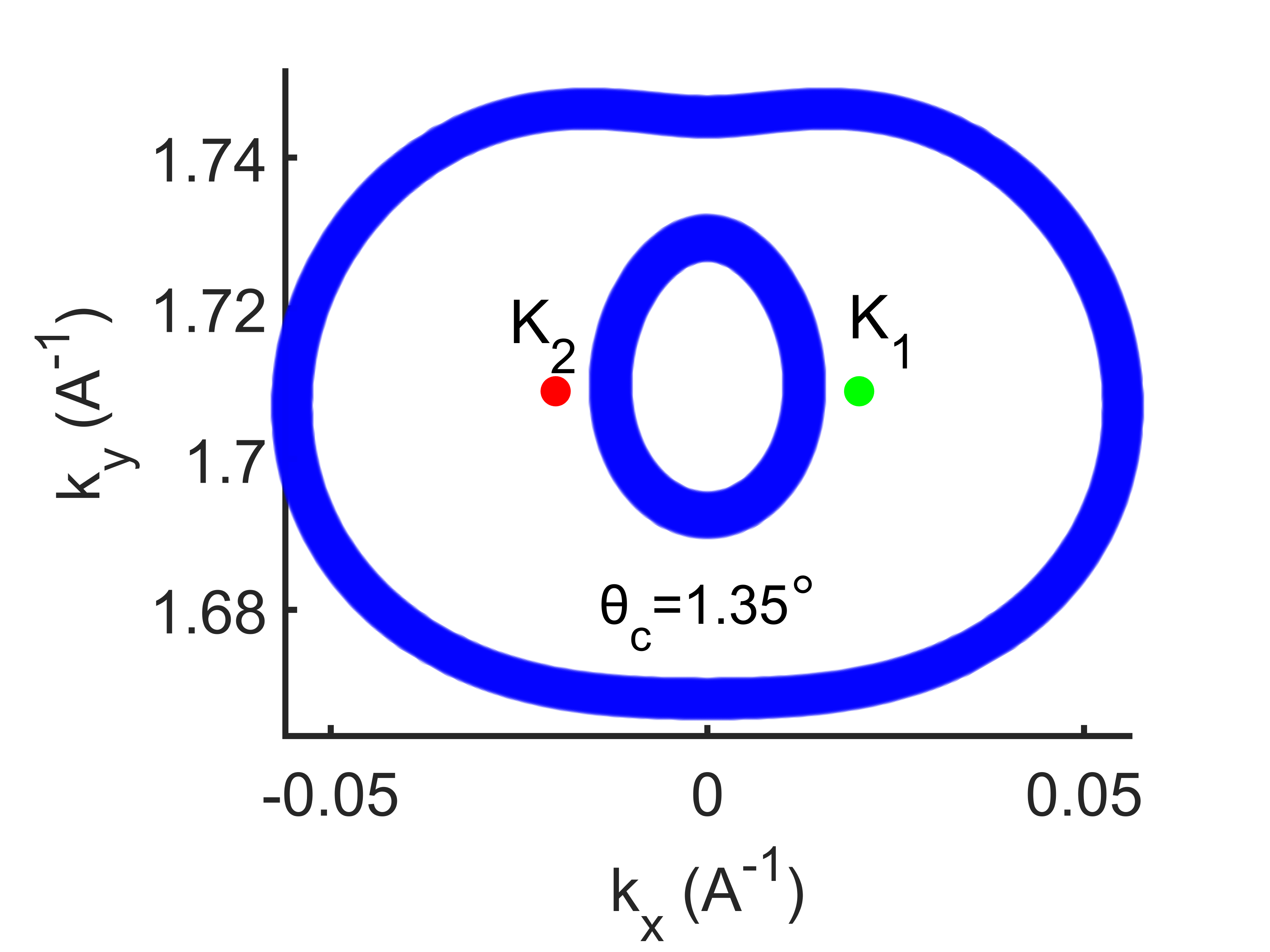}\label{Constant_energy_curve_thetaC=1.35_degree}}
	\subfloat[]{\includegraphics[width=0.33\textwidth,height=2 in]{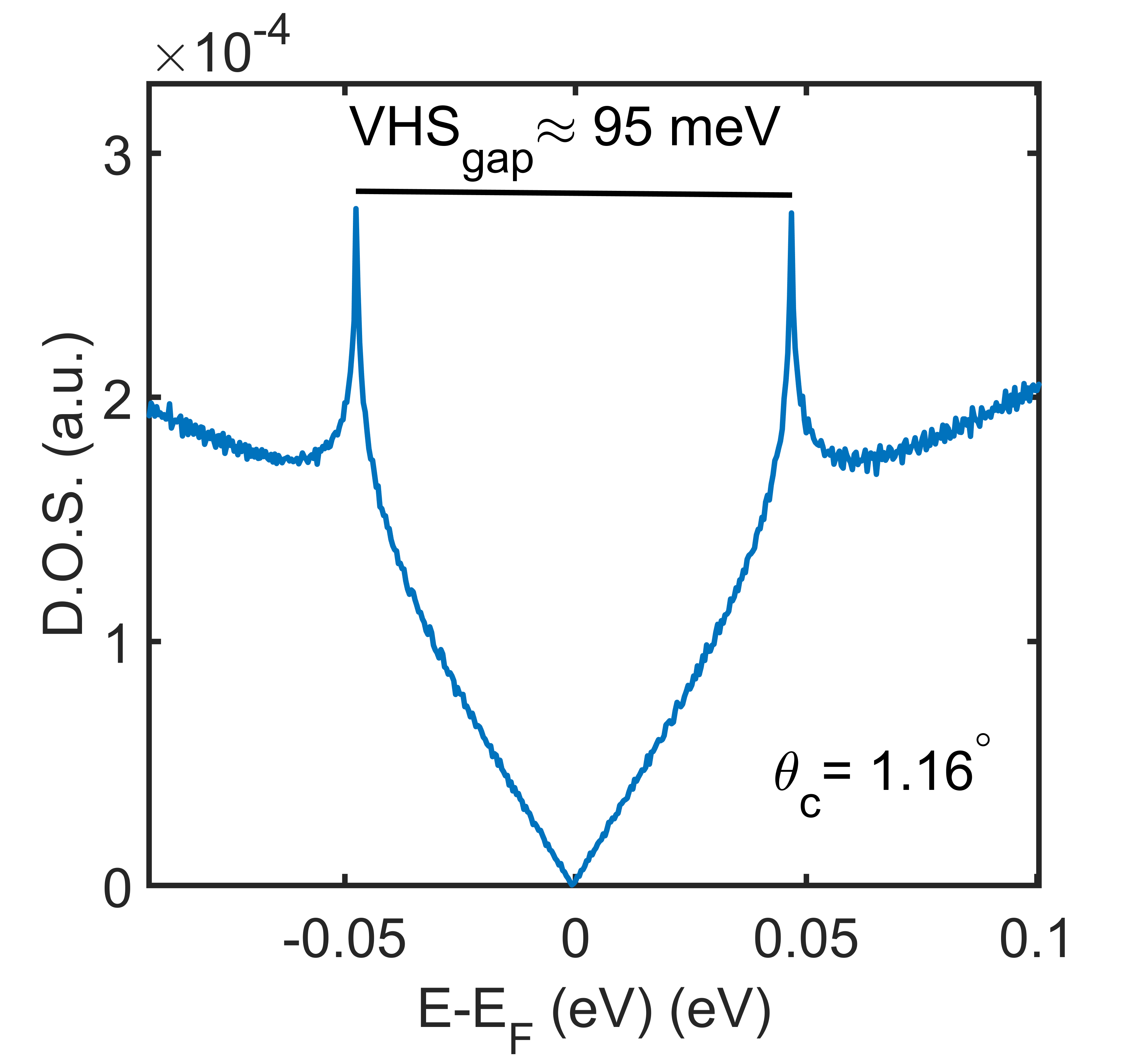}\label{TBG_VHS_gap_thetaC=1.16_degree}}
	\caption{Constant enrgy curve for TBG with twist angle (a) $\theta_c=11.64^{\circ}$, (b) $\theta_c=1.35^{\circ}$, (c) Gap between Van-Hove singularity peaks for TBG with twist angle $\theta_c=1.59^{\circ}$}\label{TBG_Final}		
\end{figure}
Figure \ref{Constant_energy_curve_thetaC=11.64_degree} shows computed constant energy contour for TBG with twist angle $\theta_c=11.64^{\circ}$ at electron energy $E_F-0.4 eV$; computed results are in good agreement with experimentally observed results presented in FIG. 1(b) of reference \cite{2012-Taisuke}. Figure \ref{Constant_energy_curve_thetaC=1.35_degree} shows computed constant energy contour for TBG with twist angle $\theta_c=1.35^{\circ}$ at electron energy $E_F-0.2 eV$; computed results differ significantly from experimentally observed results presented in Fig. 3(b) of reference \cite{2021-Lisi}. Figure \ref{TBG_VHS_gap_thetaC=1.16_degree} shows gap between two Van-Hove singularity peaks in computed D.O.S. for TBG with twist angle $\theta_c=1.16^{\circ}$. This gap between two Van-Hove singularity peaks in computed D.O.S. is $\approx 95 meV$ which differ significantly from experimentally observed gap which is $\approx 55 meV$  as reported in Fig. 2(c) of reference \cite{2019-Alexander}. Another article of reference \cite{2010-Guohong} reported this gap to be $\approx 12 meV$, indicating the inconsistency in available experimental data.\\
Computed results agree qualitatively and to a good extent quantitatively also with experimentally observed results. But for smaller twist angles this quantitative difference is significant. Theory considers ideal situation which does not take into account many experimental factors, this may cause significant difference in theoretical and experimental results. Other reason for difference between theoretical and experimental results may be absence of electronic correlations in theory which is firmly supported by scientific community \cite{2019-Yonglong, 2021-Myungchul}. Inclusion of electronic correlations in theory and better approximation of environment dependent interlayer hopping function can further enhance the results.		
\ack
We acknowledge MHRD India for providing research fellowship and Indian Institute of Technology, Roorkee for providing research facilities.
\section*{References}
\bibliographystyle{iopart-num}
\bibliography{bibfile}		
\end{document}